\documentclass{jfm}

\usepackage{amsmath} 
\usepackage{mathtools}
\usepackage{xfrac} 
\usepackage{bm} 
\usepackage{verbatim}
\usepackage{IEEEtrantools} 
	\IEEEeqnarraydefcolsep{0}{\leftmargini}
\usepackage{graphicx} 
\usepackage{caption} 
\usepackage{subcaption} 
\usepackage{placeins} 
\usepackage{epstopdf}
\usepackage{array}
\usepackage[usenames,dvipsnames]{xcolor}
\usepackage[textsize=scriptsize]{todonotes}
\usepackage{mathtools,tikz,caption}
\DeclareRobustCommand\sampleline[1]{%
  \tikz\draw[#1] (0,0) (0,\the\dimexpr\fontdimen22\textfont2\relax)
  -- (2em,\the\dimexpr\fontdimen22\textfont2\relax);%
}
\newcommand{\BV}{\textcolor{black}}
\newcommand{\RI}{\textcolor{black}}
\newcommand{\RII}{\textcolor{black}}
\newcommand{\RIII}{\textcolor{black}}

\IEEEeqnarraydefcolsep{0}{\leftmargini}

\newcommand{\mbf}[1]{\mathbf{#1}}
\newcommand{\mbs}[1]{\boldsymbol{#1}}
\newcommand{\mrm}[1]{\mathrm{#1}}

\makeatletter
\newcommand*{\getlength}[1]{\strip@pt#1}
\makeatother
\newlength{\myFigBoxWidth}
\newlength{\myFigWidth}
\newlength{\myFigHeight}
\newlength{\myFigBoxHeight}
\newlength{\xCoord}
\newlength{\yCoord}
\newlength{\myPictureHeight}
\newlength{\myHalfTextWidth}
\newlength{\myHalfBoxWidth}

\newcommand{\placeTwoSubfigures}[6]{%
\setlength{\unitlength}{1pt}%
\setlength{\myFigBoxWidth}{0.5\textwidth}%
\setlength{\myFigWidth}{\myFigBoxWidth - #4 - #5}%
\setlength{\myFigHeight}{#3\myFigWidth}%
\setlength{\myFigBoxHeight}{\myFigHeight + #6}%
\begin{picture}(\getlength{\textwidth},\getlength{\myFigBoxHeight})%
    \setlength{\xCoord}{0pt + #4}%
    \setlength{\yCoord}{-10pt}%
    \put(\getlength{\xCoord}, \getlength{\yCoord}){\includegraphics[width=\myFigWidth]{#1}}%
    \setlength{\xCoord}{\myFigBoxWidth + #4}%
    \put(\getlength{\xCoord}, \getlength{\yCoord}){\includegraphics[width=\myFigWidth]{#2}}%
    \setlength{\xCoord}{0pt}%
    \setlength{\yCoord}{\myFigBoxHeight - \heightof{\Huge{b)}}}%
    \put(\getlength{\xCoord}, \getlength{\yCoord}) {\footnotesize{(a)} }%
    \setlength{\xCoord}{\myFigBoxWidth}%
    \put(\getlength{\xCoord}, \getlength{\yCoord}) {\footnotesize{(b)} }%
\end{picture}%
}

\shorttitle{Rheology of mobile sediment beds sheared by viscous, pressure-driven flows}
\shortauthor{B. Vowinckel, E. Biegert, E. Meiburg, \'{E} Guazzelli, and P. Aussillous}

\title{Rheology of mobile sediment beds sheared by viscous, pressure-driven flows}

\author{
  B. Vowinckel\aff{1,2}
  \corresp{\email{b.vowinckel@tu-braunschweig.de}},
  E. Biegert\aff{1},
  E. Meiburg\aff{1},
  P. Aussillous\aff{3},\\
  \and
  \'{E}. Guazzelli\aff{4}  
 }

\affiliation{
\aff{1} Department of Mechanical Engineering, University of California Santa Barbara, Santa Barbara, CA 93106, USA
\aff{2} Leichtwei\ss-Institute for Hydraulic Engineering and Water Resources, Technische Universit\"at Braunschweig, 38106 Braunschweig, Germany
\aff{3} Aix Marseille Universit\'e, CNRS, IUSTI, Marseille, France
\aff{4} Universit\'e de Paris, CNRS, Mati\`ere et Syst\`emes Complexes (MSC) UMR 7057, Paris, France
}
\begin{document}

\maketitle

\begin{abstract}
We present a detailed comparison of the rheological behaviour of sheared sediment beds in a pressure-driven, \RI{straight channel configuration} based on data that was generated by means of fully coupled, grain-resolved direct numerical simulations and experimental measurements reviously published by Aussillous {\it et al.} (J. Fluid Mech., vol. 736, 2013, pp. 594-615). The highly-resolved simulation data allows to compute the stress balance of the suspension in the streamwise and vertical directions and the stress exchange  between the fluid and particle phase, which is information needed to infer the rheology, but has so far been unreachable in experiments. Applying this knowledge to the experimental and numerical data, we obtain the statistically-stationary, depth-resolved profiles of the relevant rheological quantities. The scaling behavior of rheological quantities such as the shear and normal viscosities and the effective friction coefficient are examined and compared to data coming from rheometry experiments and from widely-used rheological correlations. We show that rheological properties that have previously been inferred for annular Couette-type shear flows with neutrally buoyant particles still hold for our setup of sediment transport in a Poiseuille flow \RI{and in the dense regime we found good agreement with empirical relationships derived therefrom.}
\BV{Subdividing the total stress into parts from particle contact and hydrodynamics suggests a critical particle volume fraction of 0.3 to separate the dense from the dilute regime. In the dilute regime, i.e., the sediment transport layer, long-range hydrodynamic interactions are screened by the porous media and the effective viscosity obeys the Einstein relation. }

\end{abstract}


\section{Introduction} \label{sec:introduction}
Understanding and predicting the behaviour of mobile, granular sediment beds exposed to shearing flows is essential for a number of natural phenomena (e.\,g. sediment transport in rivers and oceans) but also for numerous engineering processes (e.\,g. slurry transport in the mining and petroleum industries). While many of these flows are turbulent, this is not always the case, and laminar flows are also present in many situations, such as the debris flow of highly concentrated suspensions or the creeping motion of soils down a hill slope \citep{jerolmack2019}. The underlying physical mechanisms leading to the morphology of sediment beds, i.e. ripples and dunes, observed in the turbulent case seem also to bear many similarities to those seen in the laminar case \citep{lajeunesse2010a}. Studies for laminar cases can thus be important to create analogues of phenomena that occur in turbulent flows at larger field scales but are also of interest by themselves. The present study focuses on this laminar regime.

Modelling sediment transport requires the understanding of the rheological behaviour of the granular material \citep[][and references therein]{vowinckel2021}. Gravity plays an important role as it controls the level of stress experienced by the grains. The motion of the grains is caused by the shearing forces exerted by the fluid at the surface of the sedimented bed but the grain packing is controlled by gravity and is free to dilate as the shearing forces are increased. This rheological situation termed pressure-imposed has been the subject of significant recent advances. It has been shown that a description in terms of a frictional rheology can be applied to both dry granular flows and viscous suspensions, despite the fact that the interactions between particles may be different. Inter-particle collisions and friction between contacting particles dominate in granular flow while hydrodynamic interactions are important in viscous suspension although the role of contacts becomes increasingly predominant with increasing concentration \citep[see e.\,g.][]{Guazzelli2018}.

In the inertial case of a dry granular material sheared at a shear rate $\dot{\gamma}$ under an imposed granular pressure $p_p$, the rheology is determined by the particle volume fraction, $\phi$, and the macroscopic friction, $\mu=\tau/p_p$, where $\tau$ is the shear stress, which both are functions of a single dimensionless inertial number $I=d_p \dot{\gamma}\sqrt{\rho_p/p_p}$, where $d_p$ is the particle diameter and $\rho_p$ the particle density \citep{GDRMidi2004,Forterre2008}. A similar formalism can be applied to viscous suspensions of non-Brownian spheres but with a viscous number $J = \eta_f \dot{\gamma}/p_p$ in place of the inertial number $I$ \citep{Boyer2011}, where $\eta_f$ is the dynamic viscosity. This frictional formulation is equivalent to the more classical presentation using viscosities depending solely on $\phi$, where the relative shear and normal viscosities can be obtained as $\eta_s=\tau/\eta_f \dot{\gamma}=\mu/J$ and $\eta_n=p_p / \eta_f \dot{\gamma}=1/J$. The transition from the viscous to the inertial regime is far from being completely understood but is supposed to occur when the Stokes number $St = I^2/J=\rho_p \dot{\gamma} d_p^2 / \eta_f$ is~$\sim O(1-10)$ \citep{bagnold1954,ness2015}.

Traditionally, the rheology of dense suspensions has been assessed in rheometry experiments of neutrally buoyant spheres, by imposing a constant volume fraction, $\phi$, to obtain the shear viscosity $\eta_s$ as a function of $\phi$ \citep[see e.\,g.][]{Stickel2005,Guazzelli2018}.  The pressure-imposed rheometry described in the preceding paragraphs is a recent addition and has been found to be particularly useful in the range of large $\phi$ which is less amenable to conventional rheometry  \citep{Boyer2011,Dagois2015,tapia2019,Guazzelli2018}. This latter approach where the suspension is free to expand (or to contract) under shear is particularly well suited to study the rheology close to the jamming transition and to measure the maximum volume fraction, $\phi_c$, where the viscosities diverge. Another valuable aspect of this pressure-controlled rheometry is the direct measurement of the particle pressure, $p_p$, which is usually not accessible to conventional rheometry.

The rheological measurements can be described by empirical correlations relating the shear viscosity, $\eta_s$, to the volume fraction, $\phi$ \citep[see e.\,g.][]{Stickel2005}. There are also phenomenological relations for the normal viscosity,  $\eta_n$,  versus $\phi$, in particular that proposed by \cite{Morris1999} to match experimental results on shear-induced migration. Of particular interest to sediment transport wherein two different phases need to be addressed are the relations proposed by \cite{Boyer2011} as the expressions for $\mu(J)$ and $\phi(J)$ and equivalently for $\eta_s(\phi)$ and $\eta_n(\phi)$ contain two terms, one coming from the hydrodynamic interactions and one coming from direct contacts. The viscous term is constructed to yield the Einstein viscosity at low $\phi$ while the contact term is similar to that found for dry granular media and produces the observed power-law divergence at the jamming transition.
\RIII{It is important to note that these phenomenological relations are inherently empirical, because they involve adjustable parameters that have been determined by best fit to experimental data of sheared dense suspensions with particle volume fractions $\phi/\phi_c>0.5$. Hence, using these correlations to describe sediment transport may be problematic, because the bed load transport layer can easily reach values lower than the data range of the rheometry experiments. Consequently, there is a need to investigate sediment transport by means of highly resolved data to test the validity of the empirical correlations as constitutive laws.} 

Modelling sediment flows on a continuum scale indeed requires applying a two-phase approach, and thus using the appropriate constitutive relationships for the stresses of the fluid and particle phases is essential. This search started with the pioneering studies of \cite{bagnold1956}, who applied the results of his rheological experiments \citep{bagnold1954} to the non-uniform case of grains flowing over a gravity bed, i.e. a sediment bed stabilized by gravity. Recognising the necessity of a frictional view of the problem has been found to be instrumental. In particular, \cite{Ouriemi2009a} used a frictional rheology similar to that proposed for dry granular media to describe the stress of the particle phase as they considered that the grains were mainly interacting through contact forces inside the bed. They also took a Newtonian rheology for the fluid phase with an Einstein dilute viscosity, as for grains in contact higher-order hydrodynamic interactions are shielded and the viscosity just reduced to its dilute value.

Two-phase modelling using a $\mu(J)$ frictional rheology has been tested against bedload experiments in channel flows by \cite{Aussillous2013} and was found to be successful in predicting the flow inside the mobile bed. However, the rheological coefficients were adjusted to match the experimental velocity and concentration profiles and overall differ from those found by \cite{Boyer2011}. Another method followed by \cite{Houssais2016} was to consider heavy particles sheared in an annular channel and to infer the rheological properties of the settled suspension. The rheology of \cite{Boyer2011} was recovered, but with the addition of a pressure at the interface of the free fluid flow and the sediment bed,  which corresponded to some fraction of the weight of an individual particle. These experiments which aim at studying the rheological behaviour of mobile sediment beds are particularly difficult as they require great accuracy in the measurements of the packing fraction and of both the particle and fluid motion. There are additional difficulties coming from the choice of the shearing flows.  In annular Couette flows \citep{Mouilleron2009,Houssais2016}, the thickness of the flowing suspension is merely two particle diameters, and using a continuum description may not be fully justified. The mobile layer is much thicker in Poiseuille-type flows and is better suited to a comparison with a continuum modelling. This setup is also a more realistic representation of a natural channel because unlike in annular flows the sidewalls have no curvature \citep{lobkovsky2008,Aussillous2013,Allen2017}. However, using a \RI{straight channel configuration} 
may also prove to be  problematic as there is a continuous erosion in the channel, and consequently the generated data always contain a transient component where erosion and deposition may not be in full equilibrium. 
 
Consequently, uncertainties remain as to how rheological models
\RIII{derived for Couette flows} 
are applicable to the situation of natural sediment transport, where 
\RIII{the flow is typically driven by a volume force, such as a pressure gradient or a downhill force. For these types of flows, the key differences are (i) total shear increasing with flow depth, which yields a wide range of viscous numbers $J$,}
(ii) settled particles where the particle pressure $p_p$ and viscous number $J$ vary vertically, and (iii) non-homogeneous particle-volume fractions throughout the bed-load transport layer at the interface between the free-fluid flow and the sediment bed. In particular, this latter region close to the interface remains poorly accessible by means of experimental measuring techniques. A promising alternative route for investigating the rheological behaviour of sediment beds is to use direct numerical simulations (DNS) at the particle scale. There are only a few contributions in the context of beload transport. In particular, the study of \cite{Kidanemariam2014}, which uses the immersed boundary method (IBM) for the fluid-solid coupling and a soft-sphere approach for solid-solid contact (DEM), investigated the flow-induced motion of a thick bed of spherical particles and found excellent agreement with the mean flow properties reported by \cite{Aussillous2013} even though the Reynolds number in the simulations was two orders of magnitude larger than in the experiment. Furthermore, \cite{Kidanemariam2017Diss} presented a first attempt to directly infer the rheology of a sheared sediment bed. The present paper is following this route to verify if previous rheological considerations from pressure-imposed rheometry remain applicable for the setup of sediment transport in a pressure-driven flow.

The objective of the present contribution is, hence, to quantify the stress exchange of the fluid and particle phase in the mixture to access  the highly-complex fluid particle interactions in the bedload transport layer. This analysis is crucial to  compute directly the rheological behavior of sediment beds exposed to a pressure-driven flow and allows for a comparison to previous rheological studies of annular Couette flows with neutrally buoyant particles. Towards this goal, we employ particle-resolved DNS using the IBM \citep{Uhlmann2005,Kempe2012a} and the DEM validated by \cite{Biegert2017a}. The simulation framework is applied to reproduce the experimental configuration of \cite{Aussillous2013}, albeit for higher flow rates but by still remaining in the viscous regime of flow with $St < 10$, see \S\,\ref{sec:experimental_data} and \ref{sec:simulations}. We then apply in \S\,\ref{sec:stress_balance} the strategy of \cite{biegert2018a} and \cite{vowinckel2019b} to capture quantities unreachable in experiments such as the stress balances for the fluid and particle phases and the whole fluid-particle mixture in the streamwise and vertical directions.  Rheological quantities are also inferred from these highly-resolved data and are compared to the reconstructed rheological data from the experimental data of \cite{Aussillous2013} in \S\,\ref{sec:rheology}. The scaling behaviour of the shear and normal viscosities as well as of the effective friction coefficient are explored and compared to the data of \cite{Boyer2011}, \cite{Dagois2015}, and \cite{tapia2019} obtained by pressure-imposed rheometry as well as with widely-used correlations \citep{Morris1999,Boyer2011}.

\section{Experimental data}\label{sec:experimental_data}

We use the experimental data of \cite{Aussillous2013} who examined the mobile layer of a granular bed in laminar flows in a rectangular-channel flow. This database yields depth-resolved profiles of particle velocity for a range of fluid heights $h_f$ as the height of the clear-liquid layer above the granular bed and flow rates in the laminar regime.


We summarise below the main features of the experimental apparatus used to obtain this database. Further details can be found in \cite{Aussillous2013}. Two batches of particles, consisting respectively of borosilicate spheres having a mean diameter $d_p=1.1\pm0.1$mm and a specific density $(\rho_p - \rho_f)/\rho_f=1.1$  and of PMMA spheres having a mean diameter $d_p=2.04\pm0.03$mm and a specific density $(\rho_p - \rho_f)/\rho_f=0.1$, were selected, where $\rho_f$ is the fluid density. 
 The particles were immersed in a viscous  fluid (mainly composed of Triton X-100 and water) having the same refractive index as the respective particles.
 A dye (Rhodamine 6G) that fluoresces when illuminated by the laser in the wavelength range greater than 555 nm was added to the fluid. The flow setup consisted of a horizontal rectangular channel of length $L_x=100$ cm, height $L_y=6.5$ cm, and  width $L_z=3.5$ cm, where $x$, $y$, and $z$ denote the streamwise, vertical and spanwise coordinate, respectively. \RII{ To create a sediment bed, the channel was filled up with monodisperse particles and fluid, then turned upside down and tilted to consolidate the sediment bed. Afterwards, the channel is set horizontally and flipped back to its original position and a pressure gradient is applied to generate a small flow rate.} This procedure was applied to fill solely the channel entrance with sedimented particles, leaving an empty buffer space near the outlet of the channel. A constant fluid flow rate, $Q_f$, was then applied to erode the sedimented bed of particles into the empty buffer space. In this way, the fluid shear stress at the top of the bed decreased as the fluid-particle interface (upstream from the buffer region) receded (i.\,e. $h_f$ increased) with time. 
\RII{The fluid-sediment interface is detected by computing the maximum change of slope of the averaged grey level profile (green line in figures \ref{fig:exp}a and c).}
Several runs were conducted for each particle type, by varying the imposed flow rate. Data for the velocity profile (particle velocity, $u_p$, and fluid velocity, $u_f$, in solely the pure fluid zone for the PMMA particles) were collected by averaging 10 images over 0.5 s every 5 s.  In addition, bulk quantities were deduced such as the time evolution of the particle bed height, $h_p$, and the flow rate per unit width, $q_{f,exp}$, by neglecting the role of the moving granular layer. The experiments were conducted in the viscous regime with a Reynolds number defined as $\Rey=\rho_f Q_f / \eta_f L_z$ ranging between $0.2 <\Rey < 1.2$. The Stokes number based on the shear rate $\dot{\gamma}$ was $St = \rho_p d_p^2 \dot{\gamma} / \eta_f \approx 0.01$ on average.

\setlength{\unitlength}{1cm}
\begin{figure}
\begin{center}
\begin{picture}(13,14)
 \put( 0,7){\includegraphics[trim=0.8cm 0 0cm 0, clip, height=7.cm]{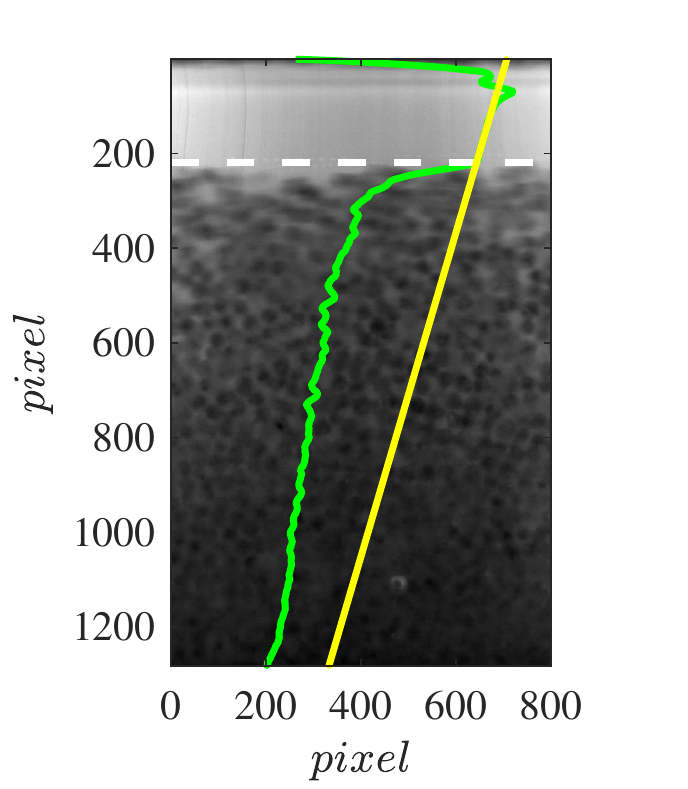}}
 \put( 5,7){\includegraphics[trim=0cm 0 0.8cm 0, clip, height=7.cm]{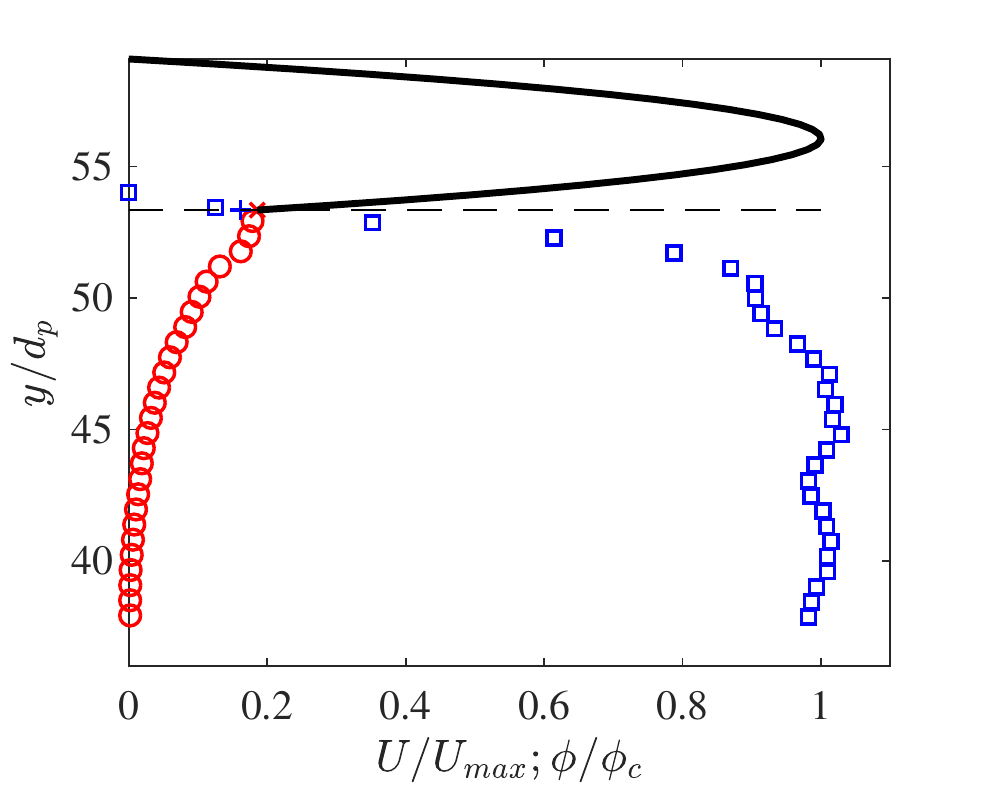}}
 \put( 0.15,0){\includegraphics[trim=0.8cm 0 0cm 0, clip, height=7.cm]{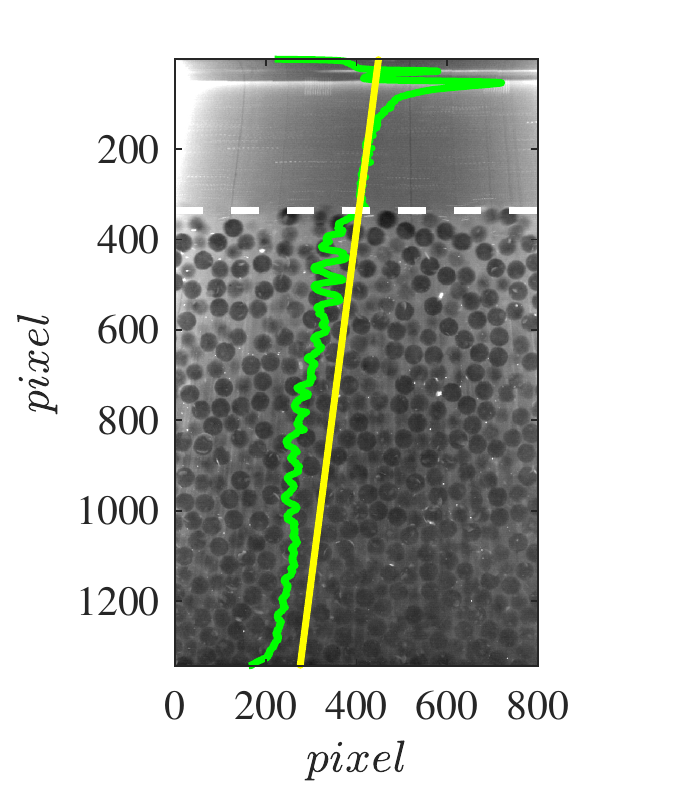}}
 \put( 5,0){\includegraphics[trim=0cm 0 0.8cm 0, clip, height=7.cm]{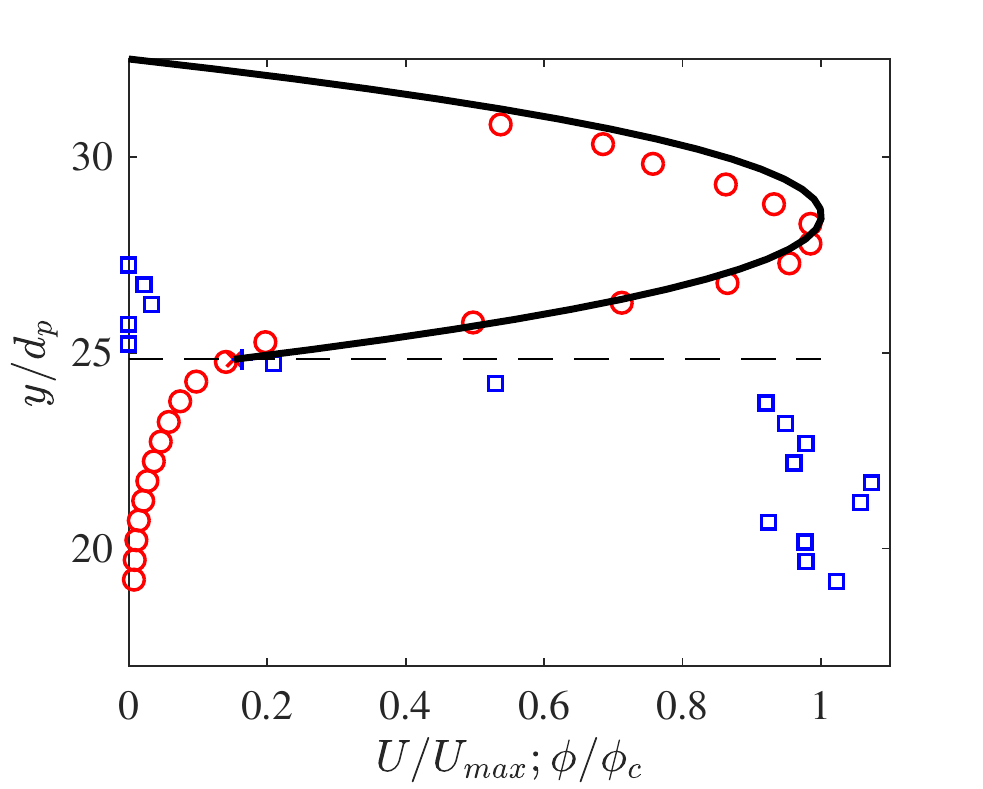}}
 \put(  0,13.5){\footnotesize{(a)}}
 \put(4.2 ,13.48){\line(1,0){0.5}}
 \put(4.45,12.98){\vector(0,1){0.5}}
 \put(4.45,12.98){\vector(0,-1){0.45}}
 \put(4.5 ,12.9 ){\footnotesize{$h_f$}}
 \put(4.2 ,12.55){\line(1,0){0.5}}  
 \put(4.45,12.05){\vector(0,1){0.5}}
 \put(4.45,12.05){\vector(0,-1){3.96}} 
 \put(4.5 ,10.0 ){\footnotesize{$h_p$}}
 \put(4.2 , 8.09){\line(1,0){0.5}} 
 \put(  5,13.5){\footnotesize{(b)}} 
 \put(  0,6.5){\footnotesize{(c)}}
 \put(4.2 ,6.48){\line(1,0){0.5}}
 \put(4.45,5.98){\vector(0,1){0.5}}
 \put(4.45,5.98){\vector(0,-1){0.88}}
 \put(4.5 ,5.7 ){\footnotesize{$h_f$}}
 \put(4.2 ,5.13 ){\line(1,0){0.5}}  
 \put(4.45,4.6){\vector(0,1){0.5}}
 \put(4.45,5.05){\vector(0,-1){3.96}} 
 \put(4.5 ,3.0 ){\footnotesize{$h_p$}}
 \put(4.2 , 1.09){\line(1,0){0.5}}  
 \put(  5,6.5){\footnotesize{(d)}}
 \end{picture}
\caption{Depth-resolved quantities extracted from the experiments for (a,b) borosilicate particles with $Q_f=2.7 10^{-6}$ m$^3$/s and  $h_f=6.3$ mm and for (c,d)  PMMA particles with $Q_f=2.7 10^{-6}$ m$^3$/s and $h_f=15.3$ mm corresponding  to runs 1 and 12 of \cite{Aussillous2013}, respectively. (a,c) Averaged images over 10 s having a length scale (a) $0.029$ mm.pixel$^{-1}$ and (b) $0.046$ mm.pixel$^{-1}$ with the corresponding grey-level profile (green curve) and the linear adjustment in the pure fluid zone (yellow straight line).  (b,d) Normalised volume-averaged velocity profile, $U/U_{max}$ (red \textcolor{red}{$\circ$}), and volume fraction profile, $\phi/\phi_c$ (blue \textcolor{blue}{$\square$}).   The full line corresponds to the theoretical velocity profile in the pure fluid zone (eq. \ref{eq:fluid_exp}). The white and grey horizontal dashed lines in graphs  (a,c) and (b,d) respectively indicate the fluid-particle interface at $y = h_p$.
}
\label{fig:exp}
\end{center}
\end{figure}

The most difficult quantity to extract from the experiments is the particle volume fraction, $\phi$. In \cite{Aussillous2013}, the volume-fraction profile for the borosilicate particles was evaluated by using the averaged grey-level profile, $P_I(y)$, of the same images as those utilised to infer the velocity profile and by scaling it by the grey-level profile of the immobile initial bed, which is assumed to be at the constant maximum particle-volume fraction $\phi_c= 0.585$. However, the laser intensity was likely to have changed during a run and the maximum packing in the bed measured with this method was  found to vary with the fluid height. We thus decide to revisit these data and to consider another method (also based on the averaged grey-level profile) to evaluate the volume fraction. As in \cite{Aussillous2013}, we suppose that the grey level intensity gradient is mainly due to a linear broadening of the laser line due to the use of a laser line generator with a finite fan angle \citep{dijksman2012}. First, we adjust the intensity decay in the pure fluid zone by a linear function, $P_I=P_{0l}(1-y/y_0)$, where  $P_{0l}$ and $y_0$ are constants which depend on the laser and the fluid properties, see figure \ref{fig:exp}(a,c) corresponding respectively to run 1 and run 12 of \cite{Aussillous2013}.  Second, for the liquid-particle mixture we assume that $P_I(y)=[(1-\phi) P_{0l}+\phi P_{0s}](1-y/y_0) $ where $P_{0s}$ is related to the particle properties. We then introduce $A=P_I(y)/(1-y/y_0)$ which is averaged over the same box size as that used for the velocity profile. If we suppose that $\phi(h_c)=\phi_c$ and $A(h_c)=A_c$ at the bottom of the mobile granular layer $h_c$, i.e. the height for which particle motion ceases,  the volume fraction is given by $\phi=\phi_c [A(y)-1]/(A_c-1)$. In figure \ref{fig:exp}(b,d),  we have plotted the normalised volume fraction, $\phi/\phi_c$, versus the vertical position made dimensionless using the particle diameter, $y/d_p$, for (b) borosilicate particles  ($Q_f=2.7 10^{-6}$ m$^3$/s and $h_f=6.3$ mm)  and (d) PMMA particles  ($Q_f=2.7 10^{-6}$ m$^3$/s and  $h_f=15.3$ mm). We observe that the volume fraction increases rapidly downward from the fluid-particle interface and reaches quickly a constant value after approximately two sphere diameters.

To examine the rheological behaviour of the sheared sediment bed, we need to infer the depth-resolved shear rate, $\dot{\mbs{\gamma}}$, the mixture shear stress, $\tau$, and the particle pressure, $p_p$, from the particle velocity profile, $u_p$, and the volume fraction profile, $\phi$. First, we interpolate linearly the velocity and volume fraction profiles to obtain their value at the fluid-bed interface, $u_{p,in}$ and $\phi_{in}$, see the red x and blue + on the dashed line in figure \ref{fig:exp}(b,d). The shear rate profile, $\dot{\mbs{\gamma}}(y)=du_p/dy$, is simply deduced numerically from the particle velocity profile using a  second-order finite difference approximation. To evaluate the total shear stress and the particle pressure, we use the two-phase modelling developed in \cite{Aussillous2013}. In this approach, a flat particle bed of thickness $h_p$ is subjected to a Poiseuille flow driven by a pressure gradient, $\partial p_f/\partial x$, in a horizontal channel. The flow is assumed to be two-dimensional, stationary, uniform in streamwise direction, parallel, and laminar.   From the fluid phase equation (i.\,e. the Brinkman equation), the volume-averaged velocity in the horizontal $x$-direction is found to be $U = \phi \, u_p + (1-\phi) \, u_f\approx u_p\approx u_f$ in the bed due to the small permeability. In the laminar regime, the momentum equations for the mixture (particles plus fluid) then write
\begin{eqnarray}
{\tau}({y})&=&{\tau}_f({h}_p)-\frac{\partial p_f}{\partial{x}}({h}_p-{y}), \label{eqqdmp2D}\\
\frac{\partial p_p}{\partial y} &=&\phi(y) \Delta\rho g , \label{eqpp}
\end{eqnarray}
where $g$ is the gravitational acceleration and $\Delta\rho=\rho_p-\rho_f$ corresponds to the density difference between the two phases.
To compute the particle pressure profile inside the bed, equation (\ref{eqpp}) is integrated numerically along the vertical $y$-direction. It shows that the pressure of the particle phase is proportional to the apparent weight of the solid phase and increases when penetrating inside the bed. The total stress profile inside the bed is deduced from equation (\ref{eqqdmp2D}), provided the stresses applied by the fluid at the fluid/bed interface, ${\tau}_f(h_p)$, and the imposed fluid pressure gradient, ${\partial {p}_f}/{\partial{x}}$, are given. 

Due to the lack of measurements of the fluid velocity in the pure fluid zone for the borosilicate particles, we need to reconstruct the fluid velocity profile for this batch of spheres based on the following assumptions:
\begin{enumerate}
\item The fluid velocity, $u_f$, is zero at the top wall ($y=L_y$) and matches the particle velocity, $u_{p,in}$, at the bed height ($y=h_p)$.
\item The fluid velocity profile is parabolic for $y>h_p$ which yields the analytical solution for a mixed Couette-Poiseuille-flow:
\begin{equation}\label{eq:fluid_exp}
u_f = \frac{{\partial {p}_f}/{\partial{x}}}{2\eta_f}\left(y^2-L_y^2\right) - \left[\frac{u_{p,in}}{h_f} + \frac{{\partial {p}_f}/{\partial{x}}}{2\eta_f}\left(h_p + L_y\right)\right](y-L_y) ,
\end{equation}
where $h_f = L_y - h_p$ is the height of the clear fluid layer .  
\item The fluid flux, $q_{f,exp}$, is written as
\begin{equation}
q_{f,exp} = \int_0^{h_p} (1-\phi) u_p \, \mrm{d}y + \underbrace{\int_{h_p}^{L_y} u_f \, \mrm{d}y}_{q_f} ,
\label{qf}
\end{equation}
considering that the fluid velocity matches the particle velocity inside the bed.
\end{enumerate}
Using the experimental flow rate per unit width, $q_{f,exp}$, and the experimental  vertical profiles of $\phi$ and $u_p$, the pure fluid flow rate $q_f$ is deduced from equation (\ref{qf}) by numerical integration and the pressure gradient for the parabolic velocity profile is then obtained as ${\partial {p}_f}/{\partial{x}} = ({6 \eta_f}/{h_f^3}) \left(h_f u_{p,in} - 2 q_f\right)$.  Inserting this value in equation (\ref{eq:fluid_exp}), we can calculate the fluid velocity profile in the pure fluid zone, see the black full curves in figures \ref{fig:exp}(b-d). The agreement with the experimental fluid velocity measurement for the PMMA particles is found to be good, see figure \ref{fig:exp}(d). This validates the present reconstruction method and its use for the borosilicate particles. Note that this method does not enforce a continuity of stress from the clear fluid phase to the particle phase; the obtained fluid and particle velocity profiles exhibit a change in slope at the fluid/bed interface.  The fluid shear stress at the bed interface is then given by $\tau_f(h_p)=  \frac{1}{2}\frac{\partial {p}_f}{\partial{x}}\left(L_y-h_p \right)- \eta_f u_{p,in}/h_f$ and the vertical profile of the shear rate is simply given by equation (\ref{eqqdmp2D}).

The extraction and reconstruction methods described above provide all the quantities needed to investigate the rheology of the mobile sediment bed: $\mu(J)$ and $\phi(J)$ or equivalently $\eta_s(\phi)$ and $\eta_n(\phi)$. 

\section{Simulation data}\label{sec:simulations}
In the present work, we use the framework described in detail in \citet{Biegert2017a,Biegert2017b} to execute several simulations in an attempt to compare to the different experimental results of \cite{Aussillous2013} at different flow rates and fluid  heights. In order to keep the paper self-contained, we provide a brief summary of the computational approach. 

The particle-laden flows of interest require us to solve the Navier-Stokes equation
\begin{equation} \label{eq:NS}
\rho_f \left[\frac{\partial{\mbs{u}_f}}{\partial{t}}+\nabla\cdot(\mbs{u}_f\mbs{u}_f)\right] = \nabla \mbs{\tau}_f + \mbs{f}_b + \mbs{f}_\mathit{IBM} \hspace{0.5cm},
\end{equation}
where $\mbs{u}_f=(u_f,v_f,w_f)^T$ denotes the fluid velocity vector and $t$ time. The fluid stress tensor is given by $\mbs{\tau}_f = -p_f \mbs{E} + \eta_f [\nabla \mbs{u}_f + (\nabla \mbs{u}_f)^T]$, where $p_f$ represents the fluid pressure with the hydrostatic component subtracted out and $\mbs{E}$ the identity matrix. The right-hand side includes the volume forces $\mbs{f}_b=(f_{b,x},0,0)^T$ and $\mbs{f}_\mathit{IBM}$. The former is a source term used to create the pressure gradient driving the flow and the latter an immersed boundary force used to enforce the no-slip condition on the particle surface. We discretize the equations of motion for the fluid on a cubic finite difference mesh ($\lambda = \Delta x = \Delta y = \Delta z$).

The numerical treatment is based on the IBM for fluid-particle coupling \citep{Uhlmann2005,Kempe2012a} and the scheme of \citet{Biegert2017a} for particle-particle interaction. 

We solve for the translational velocity, $\mbs{u}_p$,
\begin{equation} \label{eq:p_translational}
 m_p\: \frac{\text{d}\mbs{u}_p}{\text{d} t} = \underbrace{\oint_{\Gamma_p} \mbs{\tau}_f \cdot \mbs{n}\: {\text{d}A}}_{=\mbs{F}_{h,p}} +
        \underbrace{V_p\:( \rho_p-\rho_f )\: \mbs{g}}_{=\mbs{F}_{g,p}} + \mbs{F}_{i,p} \qquad ,
\end{equation}
and the angular velocity, $\mbs{\omega}_p$,
\begin{equation} \label{eq:p_rotational}
I_p \:\frac{ \text{d}\boldsymbol{\omega}_p}{\text{d} t} = \underbrace{\oint_{\Gamma_p} \mbs{r}_p\times(\boldsymbol{\tau}_f\cdot\mbs{n})\:{\text{d}A}}_{=\mbs{T}_{h,p}} + \mbs{T}_{i,p} \hspace{0.5cm},
\end{equation}
of spherical particles, where $m_p$ is the particle mass, $I_p$ the particle moment of inertia, $V_p$ the particle volume,  and $\mbs{g}=(0, -g, 0)^T$ the gravitational acceleration vector. The fluid acts on the particles through the hydrodynamic stress tensor $\mbs{\tau}_f$, where $\mbs{r}_p$ represents the vector from the particle center to a point on the surface $\Gamma_p$ and $\mbs{n}$ is the unit normal vector pointing outwards from that point. The net force and torque acting on the particle center of mass due 
\BV{to particle interactions are given by $\mbs{F}_{i,p}$ and $\mbs{T}_{i,p}$,
} 
respectively.

We evaluate the 
\BV{particle interaction}
forces and torques according to \citet{Biegert2017a}. The resulting collision model involves normal contact forces, $\mbs{F}_{n}$, tangential (frictional) contact forces, $\mbs{F}_{t}$, and short-range 
\BV{
hydrodynamic
}
lubrication forces, $\mbs{F}_{l}$, to provide the total collision force
\begin{equation}\label{eq:particle_forces}
	\mbs{F}_{i,p} = \sum_{q,\:q \neq p}^{N_{tot}} \left( \mbs{F}_{l,pq} + \mbs{F}_{n,pq} + \mbs{F}_{t,pq}\right) \hspace{0.5cm},
\end{equation}
where $N_{tot}$ is the total number of particles and the subscript $pq$ indicates interactions of particle $p$ with particle $q$. According to \cite{cox1967}, we model the unresolved 
\BV{
hydrodynamic
}
component of the lubrication forces in our simulations as
\begin{equation}\label{eq:lubrication}
  \mbs{F}_{l,pq} = 
  \begin{cases}
  - \frac{6 \pi \eta_f R_\text{eff}^2}{\max(\zeta_n,\zeta_\text{min})} \mbs{g}_n  & 0 < \zeta_n \leq 2h \\
    0 & \text{otherwise}
  \end{cases}
\end{equation} 
where $\zeta_n$ is the gap size, $R_\text{eff}=R_p R_q/(R_p + R_q)$ the effective radius, and $R_p$ and $R_q$ are the radii of the two interacting particles. Furthermore, $\zeta_\text{min}=3\cdot10^{-3}R_p$ is a limiter  as calibrated by \citet{Biegert2017a} that can be interpreted as the roughness of the particle  surface and $\mbs{g}_n$  is the normal component of the relative velocity of the two colliding particles. The repulsive normal component is represented by a nonlinear spring-dashpot model for the normal direction
	\begin{equation} \label{eq:acm_force}
	\mbs{F}_{n,pq} = -k_n |\zeta_n|^{3/2} \mbs{n} - d_n \mbs{g}_{n} \qquad ,
	\end{equation}
where $k_n$ and $d_n$ represent stiffness and damping coefficients that are adaptively calibrated for every collision/contact to prescribe a restitution coefficient of ${e_\text{dry}=-u_\text{o}/u_\text{i}=0.97}$ \citep{kempe2012b}. Here, $u_\text{o}$ and $u_\text{i}$ indicate the normal components of the relative particle speed immediately after and right before the particle contact, i.e. $\zeta_n=0$. The forces in the tangential direction are modeled by a linear spring-dashpot system capped by the Coulomb friction law as
\begin{equation} \label{eq:lin_tan}
\mbs{F}_{t,pq} = \min \left(-k_t  \boldsymbol{\zeta}_t - d_t \mbs{g}_{t,cp} , ||\mu_f \mbs{F}_n|| \mbs{t} \right)  \qquad ,
\end{equation}
where $k_t$ and $d_t$ are stiffness and damping computed according to \cite{thornton2011} and \cite{thornton2013}, $\mu_f=0.15$ represents the friction coefficient between the two surfaces, $\boldsymbol{\zeta}_t$ is the tangential displacement integrated over the time interval for which the two particles are in contact \citep{thornton2013} and $\mbs{t}$ is a unit vector pointing into the tangential direction.  The empirical parameters $e_\text{dry}=0.97$ and $\mu_f=0.15$ have been taken from experiments involving glass spheres \citep{Gondret2002,Joseph2004}. Using these values, the contact-model has been validated in detail by \cite{Biegert2017a} against seminal experimental benchmark data involving the same material \citep{Foerster1994,Gondret2002,TenCate2004,Aussillous2013}. Note that these parameters were not measured for the particles used in the experiments of \cite{Aussillous2013} and may, hence, be different.

\begin{figure}
\placeTwoSubfigures{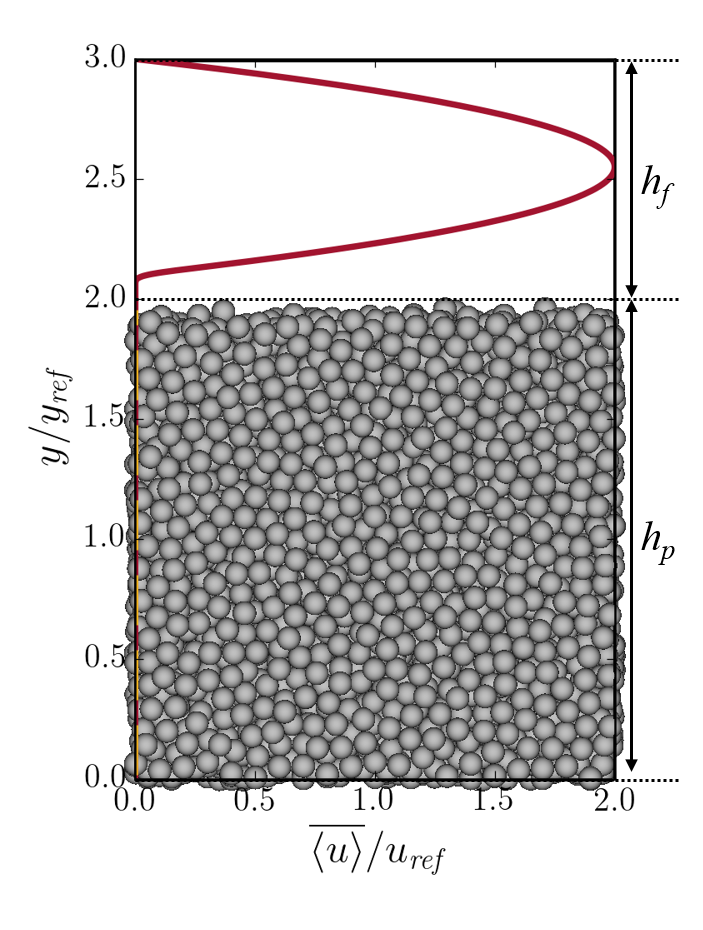}{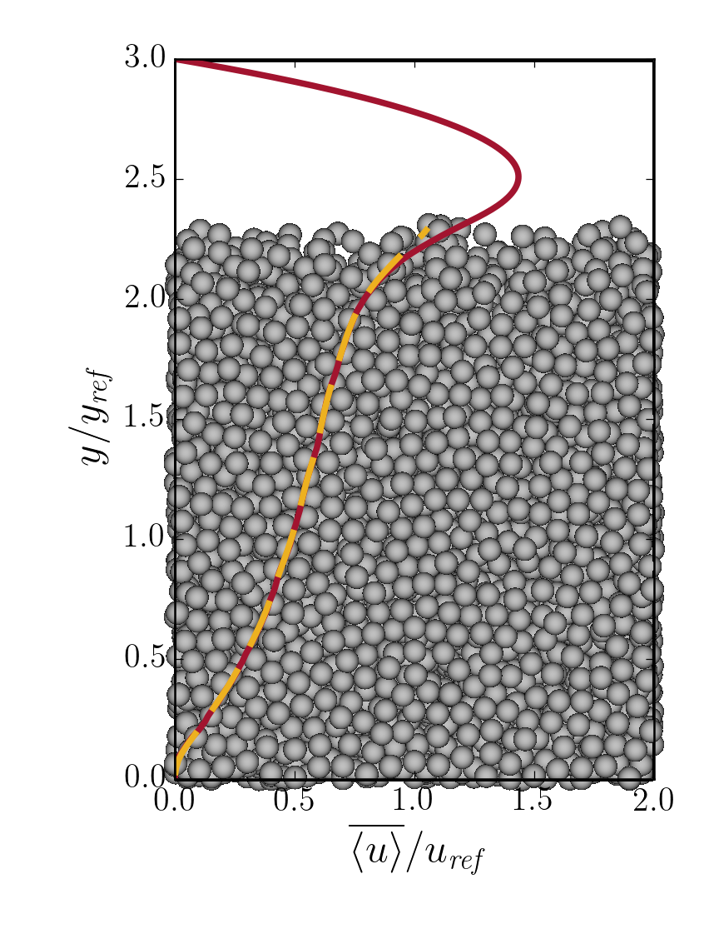}{1.3}{0pt}{10pt}{0pt}
\caption{(a) Initial conditions for fluid velocity and bed configuration at $t/t_\mathit{base}=0$ and (b) fully developed state at $t/t_\mathit{base}=10$ for simulation Re67 as listed in table~\ref{tab:bed_runs}.  Lines indicate the average streamwise ($x$-direction) fluid velocity (dark gray, red online) and particle velocity (light gray, yellow online), where the consecutive averaging in the horizontal plane and time is defined by \eqref{eq:averaging_operator}.}
\label{fig:bed_start_end}
\end{figure}

\begin{table}
\centering
\begin{tabular}{ll}
Number of particles $N_\mathit{tot}$   & 4339 \\
Galileo number $Ga=\frac{\rho_f}{\eta_f}\sqrt{\left(\frac{\rho_p}{\rho_f}-1\right)\,g\,d^3_p }$				& 0.85	\\
$\rho_p / \rho_f$	& 2.1	\\
Timestep		& $\mrm{CFL}=0.5$ \\
Domain size ($L_x/d_p \times L_y/d_p \times L_z/d_p$)
	& $20 \times 30 \times 10$	\\
Domain grid size ($L_x/\lambda \times L_y/\lambda \times L_z/\lambda$)
	& $512 \times 768 \times 256$	\\
Domain boundary conditions
	& periodic $\times$ no-slip $\times$ periodic	\\
Initial $h_f / d_p$	& 10.0	\\
Particle resolution, $d_p/\lambda$	& 25.6	\\
\end{tabular}
\caption{Simulation parameters for the pressure-driven flow over a bed of particles.}
\label{tab:bed_parameters}
\end{table}

\begin{table}
\centering
\begin{tabular}{lccccc}
Simulation run	& $Re_\mathit{ref}$	& $Sh_\mathit{ref}$ &  $t_\mathit{sim}/t_\mathit{base}$	& $t_\mathit{avg}/t_\mathit{base}$ \\
Re67	& 66.7	& 11.1	  & $[0.00,10.00]$	& initialization \\
Re17	& 16.7	& 5.53	  & $[10.00,47.20]$	& $[16.00,47.20]$ \\
Re33	& 33.3	& 2.77	  &$[10.00,58.80]$	& $[44.00,52.15]$ \\
Re8	    & 8.33	& 1.38	  &$[47.20,92.05]$	& $[77.00,92.05]$
\end{tabular}
\caption{Simulation parameters for different runs of the pressure-driven flow over a bed of particles. The Reynolds and Shields numbers based on the reference case (predefined Poiseuille flow above the sediment bed), i.e. $Re_\mathit{ref} = \rho_f u_\mathit{ref} y_\mathit{ref} / \eta_f$ and $Sh_\mathit{ref} = \sigma_\mathit{ref}/\left[(\rho_p-\rho_f) g d_p\right]$, respectively.  The individual simulation is run for the duration $t_\mathit{sim}$, and the momentum balance is analyzed by using time-averaged data over the interval $t_\mathit{avg}$. 
}
\label{tab:bed_runs}
\end{table}

In the present work, we consider a computational setup very similar to the one presented in \citet{Aussillous2013} (figure \ref{fig:bed_start_end}a). The domain has dimensions ${L_x \times L_y \times L_z = 20d_p \times 30d_p \times 10d_p}$. We discretize the domain with a regular grid that is equidistant in all three directions and has a resolution of 25.6 grid cells per particle diameter,
\RII{which is a fine enough resolution to obtain results independent of the grid cell size as shown by \cite{biegert2018a}.} 
\RII{We generate the bed by allowing 4,339 monodisperse particles to settle under gravity, without the influence of the surrounding fluid, onto a layer of 200 fixed particles whose centers randomly vary in height above the bottom wall within a range of $d_p$, providing an irregular roughness \citep{Jain2017}. The resulting bed fills the domain to about a height of $h_p \approx 20d_p$ from the bottom wall, where $h_p$ is the particle bed height, leaving a gap of about $10d_p$ between the top wall and the top of the particle bed.}
For the simulation runs, we use a particle density $\rho_p/\rho_f=2.1$ and a Galileo number of $Ga=0.85$. A definition of $Ga$ as well as a summary of the relevant simulation parameters is given in table \ref{tab:bed_parameters}. We employ a predefined Poiseuille flow in the clear fluid region above the sediment bed as a reference case for the simulation. For this reference case, we define the reference length, $y_\mathit{ref} = 10d_p = L_y/3$, the reference velocity of $u_\mathit{ref} = -y_\mathit{ref}^2 f_{b,x} / (12 \eta_f)$, and the reference stress, $\sigma_\mathit{ref} = -y_\mathit{ref} f_{b,x} /2$ to compute the corresponding Reynolds and Shields numbers (cf. table \ref{tab:bed_runs}).

\begin{figure}
\placeTwoSubfigures{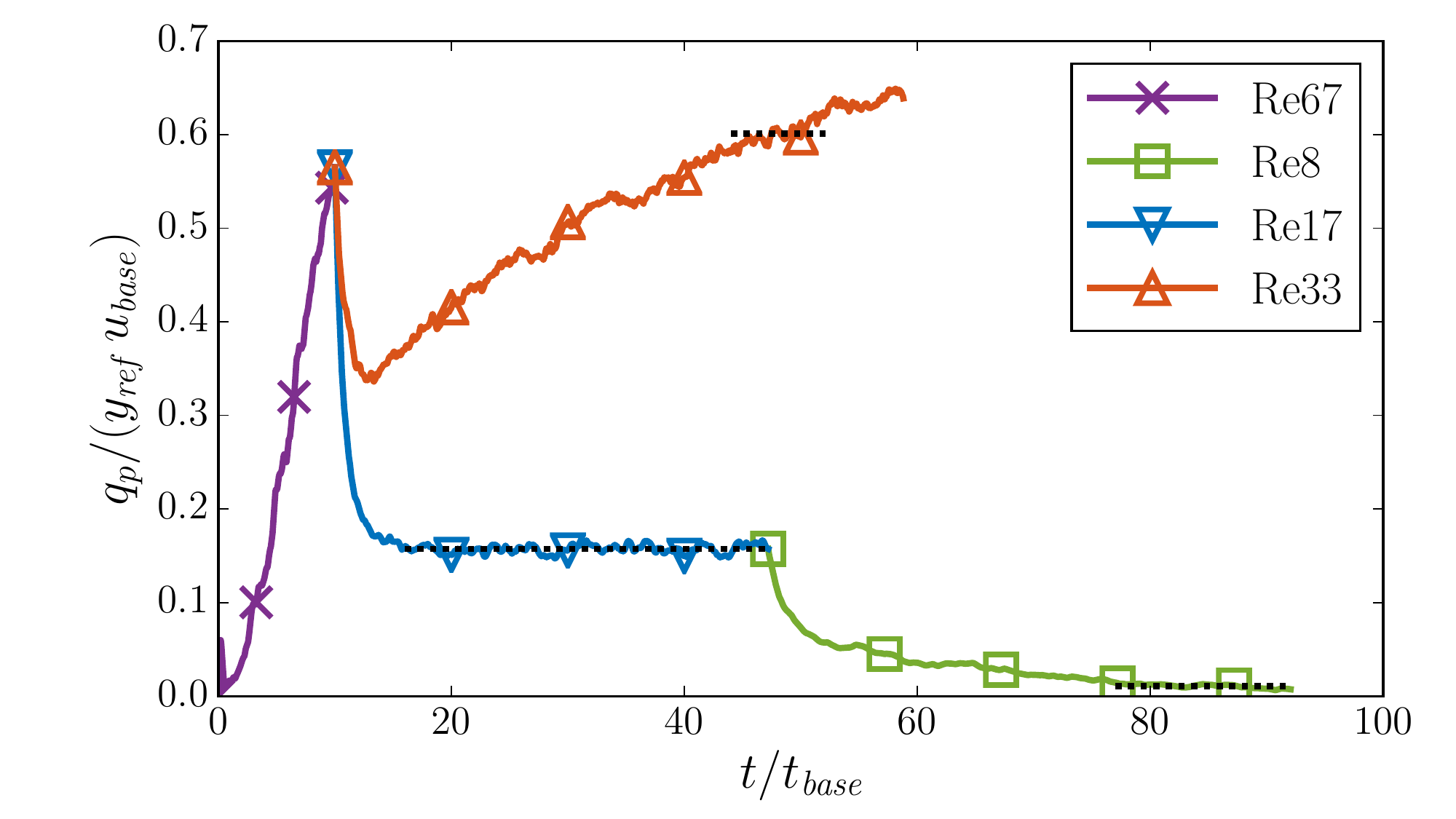}{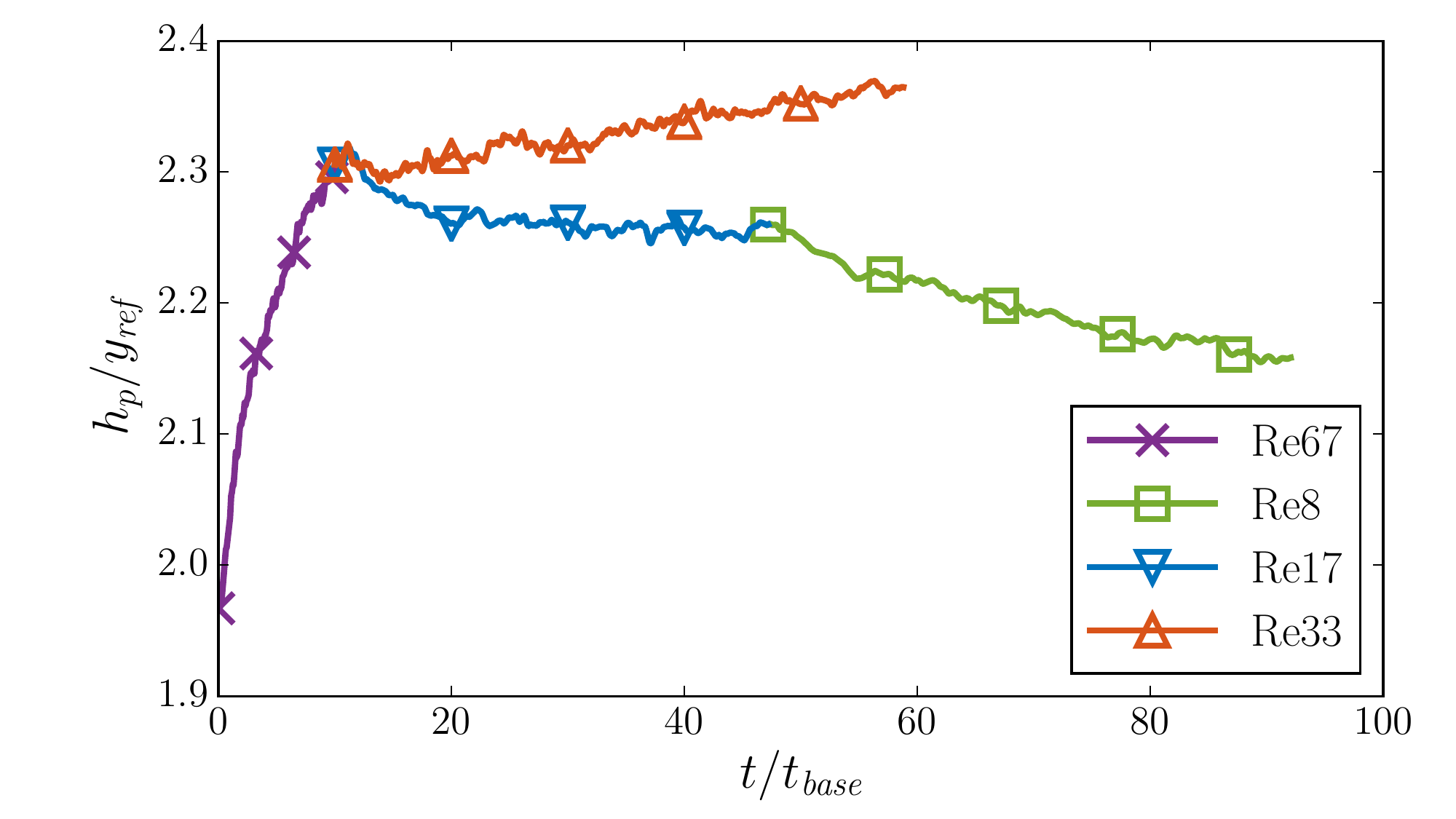}{0.55}{0pt}{0pt}{0pt}
\caption{Sheared particle bed: (a) Particle volumetric flux $q_p = \frac{1}{L_x \, L_z} \sum_{p=1}^{N_\mathit{tot}} V_p u_{p,x}$, for the different simulation runs.  Dotted lines indicate the average particle flux over the averaging time interval for each simulation. (b) Bed height, defined at $\left<\phi\right> = 0.05$.
}
\label{fig:bed_p_flux}
\end{figure}

We enforce different volumetric flow rates, governed by the volume force $f_{b,x}$. As it can take a long time for a simulation to reach a statistically steady state when initialized from rest, we found it to be more efficient to obtain a steady state by initializing the flow with a large pressure gradient that mobilizes the entire bed (run Re67 in figure~\ref{fig:bed_p_flux}a and table~\ref{tab:bed_runs}). To allow for a direct comparison among the different runs, we therefore define $u_{base}=u_{ref} (\text{Re67})$, $t_{base} = y_{ref} / (1.5 u_{base})$ and $\sigma_{base} = \sigma_{ref}(\text{Re67})$. By the end of the simulation Re67, the bed has dilated to a height of $h_p/y_\mathit{ref} \approx 2.3$ (figure \ref{fig:bed_p_flux}b), and the particles just above the fixed layer at the bottom of the domain are moving (figure~\ref{fig:bed_start_end}b). After this initialization phase, the imposed pressure gradient is reduced to produce simulations Re17 and Re33. Re8 is carried out by continuing Re17 with an even lower imposed pressure gradient. This procedure allows us to quickly reach a steady state for runs Re17 and Re8, as can be seen in figure \ref{fig:bed_p_flux}a, whereas Re33 is still in the process of dilation.

To compare the simulation results to rheological models, we have to analyze the discrete information of the particle-phase in our simulations, such as particle velocities and forces, from a continuum viewpoint. To this end, we employ the coarse-graining method (CGM) based on the works of \citet{Goldhirsch2010} and \citet{Weinhart2012}.  This coarse-graining method conserves quantities of interest, but additionally smoothes out the resulting continuum field. For a given discrete particle quantity $\theta_p$, we define its coarse-grained continuous counterpart, $\theta^\mathit{cg}$, as
\begin{equation} \label{eq:cg_general}
\theta^\mathit{cg}(\mbs{x},t) = \sum_{p=1}^{N_\mathit{tot}} \theta_p \mathcal{W}(\mbs{x} - \mbs{x}_p(t)) ,
\end{equation}
where $\mbs{x}_p(t)$ is the position of the center of particle $p$, and $\mathcal{W}(\mbs{r})$ is the conservative coarse-graining function that smears a given local quantity in a spherical volume of radius $|\mbs{r}|$. The main property is that $\int_{\mathbb{R}^3} \mathcal{W}(\mbs{r}) \, \text{d}\mbs{r} = 1$. We implemented a coarse-graining function based on the Dirac delta function of \cite{Roma1999}, which is the same delta-function used in the IBM \citep{Uhlmann2005}. The filter $|\mbs{r}|$ has to be as small as possible to fully exploit the highly-resolved simulation data for our rheological analysis, but large enough to smooth out the sub-particle scale. Here, we choose $\lvert \mbs{r}\rvert = 1.5 d_p$, which was deemed to be a good compromise to satisfy these requirements.

The steady-state configuration for the moving bed is steady only in a time-averaged sense, because particle collisions and positions continuously fluctuate.  We therefore define the time average of a given continuous field $\theta$ to be
\begin{equation} \label{eq:averaging_operator}
\overline{\left< \theta\right>}(y) = \frac{1}{t_\mathit{avg,2}-t_\mathit{avg,1}} \int_{t_\mathit{avg,1}}^{t_\mathit{avg,2}}\frac{1}{L_x L_z} \int_0^{L_z} \int_0^{L_x}   \theta(x,y,z,t) \, \mrm{d}x \, \mrm{d}z \, \mrm{d}t ,
\end{equation}
%
where the overbar and angular brackets represent averaging in time and space, respectively \citep{vowinckel2014,vowinckel2017a,vowinckel2017b}. 
\RII{Note that this averaging operator applies for both, continuous fluid quantities and coarse-grained particle quantities.}
We present the values for $t_\mathit{avg,1}$ and $t_\mathit{avg,2}$ in table~\ref{tab:bed_runs}. These time-averaging windows were chosen to capture the steady-state results, or as large a time span as possible for as similar a particle flux as possible (figure \ref{fig:bed_p_flux}). We will show, however, that our results for the rheology are independent of transient behavior for all Reynolds numbers investigated. 

\begin{figure}
\centering
\placeTwoSubfigures{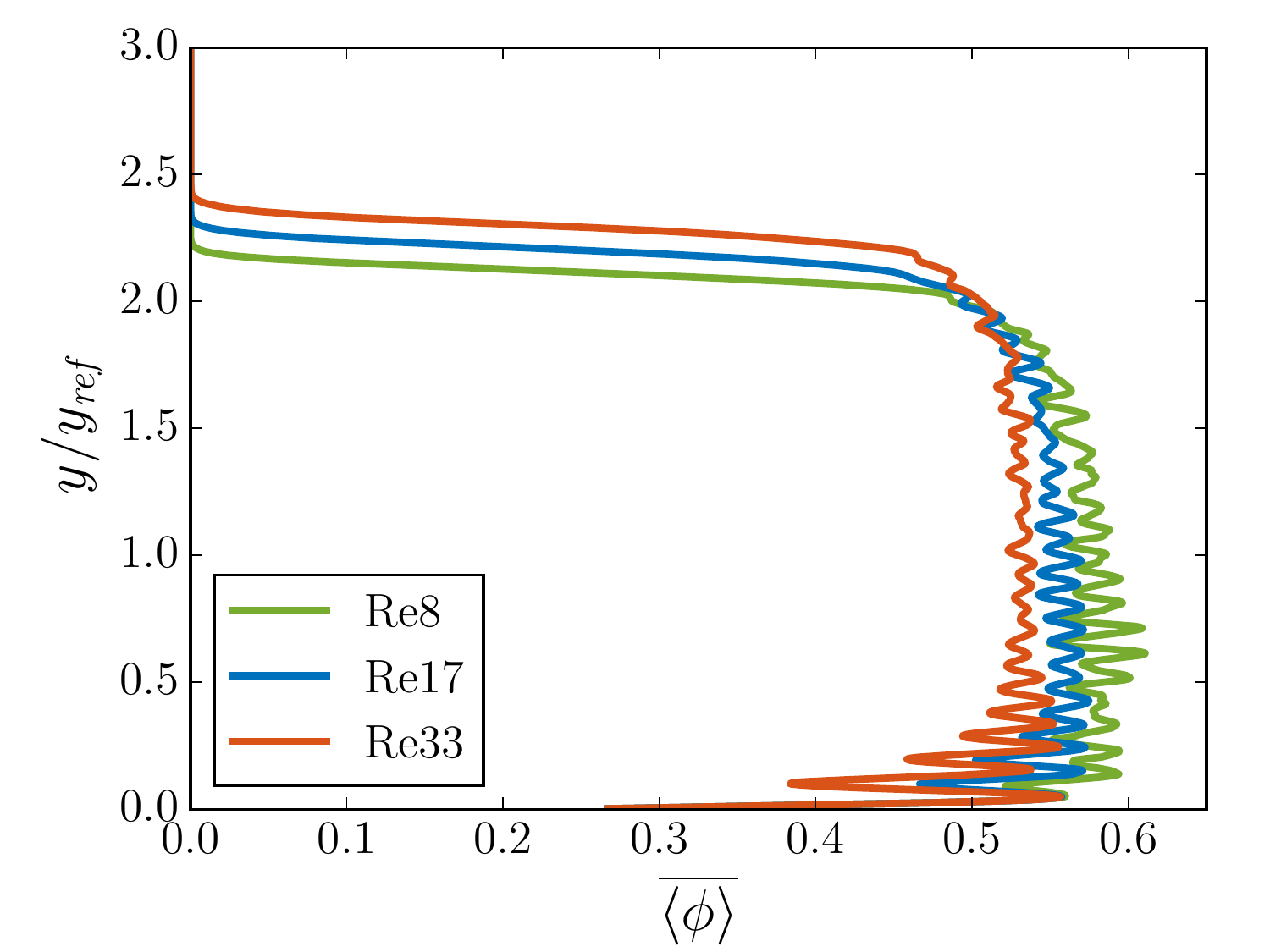}{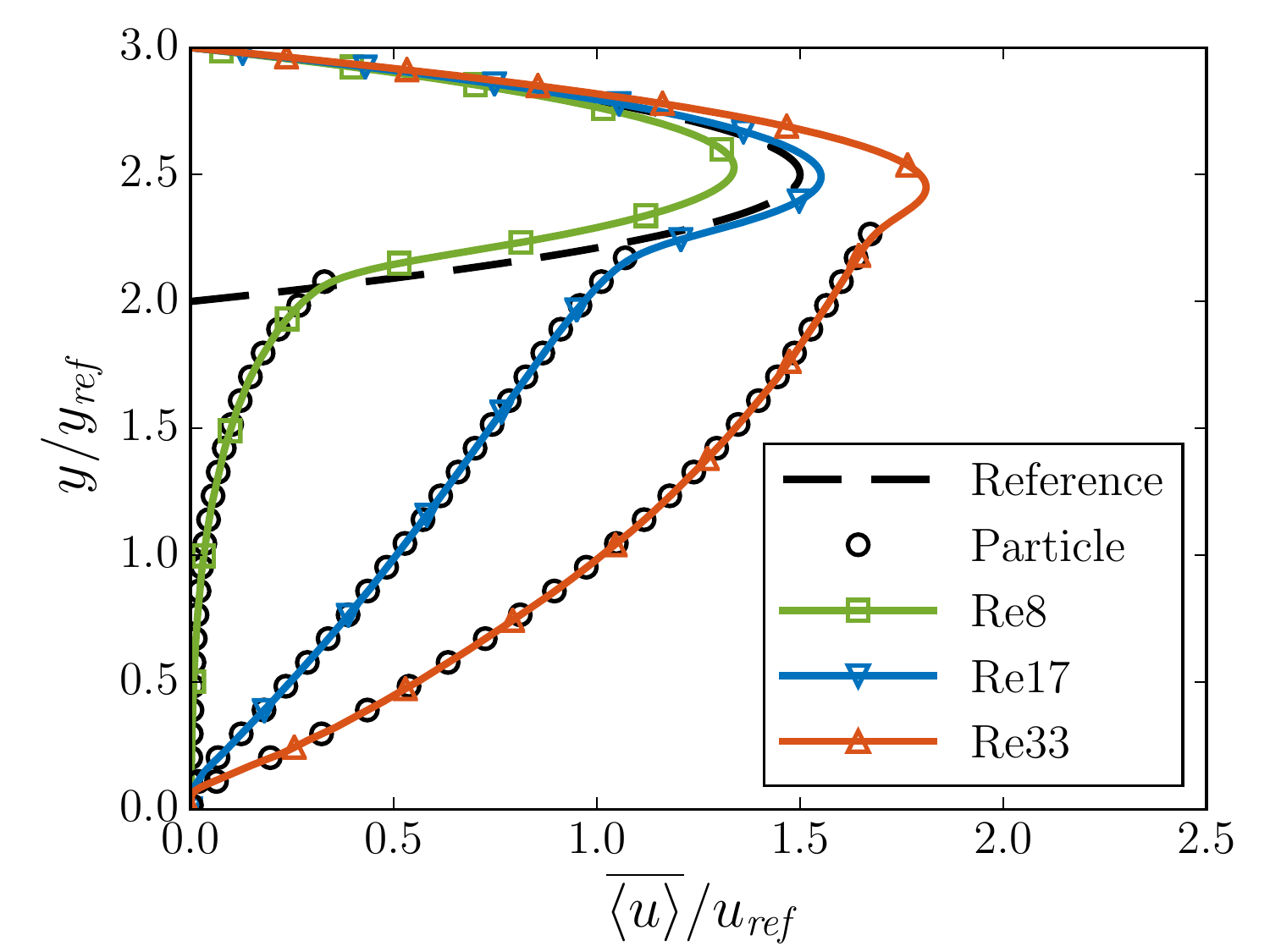}{0.75}{0pt}{0pt}{0pt}
\caption{Sheared particle bed: profiles for the different simulation runs averaged horizontally and in time for (a) the particle volume fraction and (b) the streamwise velocity.  The average fluid velocity (solid colored lines) is given by \eqref{eq:averaging_operator}, while the average coarse-grained particle velocity (circles) is given by \eqref{eq:cg_general}.
}  \label{fig:bed_flow}
\end{figure}

We can use \eqref{eq:cg_general} to obtain a continuous field of $\phi$ and  \eqref{eq:averaging_operator} to generate vertical profiles  of $\overline{\left<\phi\right>}$ and $\overline{\left<u_f\right>}$ by computing double-averaged values (in space and time, cf. figure~\ref{fig:bed_flow}). For the analysis of the rheology, we exclude values for $y/y_\text{ref}<0.68$ to eliminate boundary effects from the artificial bed roughness at the bottom all.  We observe the same evolution for the volume fraction as in the experiments with a rapid increase downward from the fluid-particle interface to reach a constant value after a few sphere diameters. 
\RII{Here, we follow the definition of \cite{Biegert2017a} to determine the fluid-sediment interface at $\overline{\left<\phi\right>}=0.05$.}
The oscillations of $\overline{\left<\phi\right>}$ in this region reflect the sub-particle scale as the particle layering within the sediment bed (figure~\ref{fig:bed_flow}a).
\RII{
We, therefore apply the coarse-graining radius of $|\mbs{r}|= 1.5 d_p$ to the grid resolved data shown in this figure to smooth out the layered structure in these vertical profiles.
}
Note that figure \ref{fig:bed_flow}b yields $\dot{\gamma}=\partial \overline{\left<u_f\right>}/\partial y$ as another rheological quantity. Looking at the double-averaged fluid velocity profiles in figure \ref{fig:bed_flow}b, the three cases represent three distinctively different regimes \citep{Jenkins2017}. The sediment bed of Re8 reaches a quasi-static regime of the granular suspension, whereas the sediment motion in Re17 and Re33 can be considered ``layered" and ``collisional", respectively. There is a clear qualitative difference between run Re8, whose velocity profile is concave and goes to zero within the bed at $y/y_\mathit{ref} \approx 0.5$, and run Re33, whose velocity profile is convex and goes to zero only at the fixed particles at the lower wall. Note that the fluid velocity is to a good approximation equal to the particle velocity, which is consistent with the observation of \cite{Aussillous2013}. The Stokes number ranges from $0.19\leq St\leq0.77$, which is 20 to 80 times larger than the values obtained from the experimental data \S \ref{sec:experimental_data}, but still expected to be in the viscous regime \citep{ness2015}. 

While we can readily extract the quantities for $\phi$ and $\dot{\mbs{\gamma}}$ from figure \ref{fig:bed_flow}, an additional investigation of the total stress balance of the fluid-particle mixture in $x$- and $y$- direction is needed to compute the total shear stress $\tau$ and the granular pressure $p_p$ as will be detailed in the next section.

\section{Stress balance of the simulation data}\label{sec:stress_balance}
\RI{This section presents the analysis of the stress balance for the fluid-particle mixture to compute wall-normal profiles of the rheological quantities  $\tau$ and $p_p$. To this end, we follow the argument of the  previous numerical work \citep{biegert2018a,biegert2018b,vowinckel2019b}, where the full derivation of the stress balances in both shearing ($x$) and wall-normal ($y$) direction are given from first principles. In addition, a detailed analysis of the present datasets for all components entering the stress balances in shear and wall-normal direction, respectively, can be found in \cite{biegert2018b}.}

\subsection{Stress balance in the $x$-direction to obtain the total stress}\label{sec:total_stress}
\begin{figure}
\centering
\includegraphics[width=0.5\textwidth]{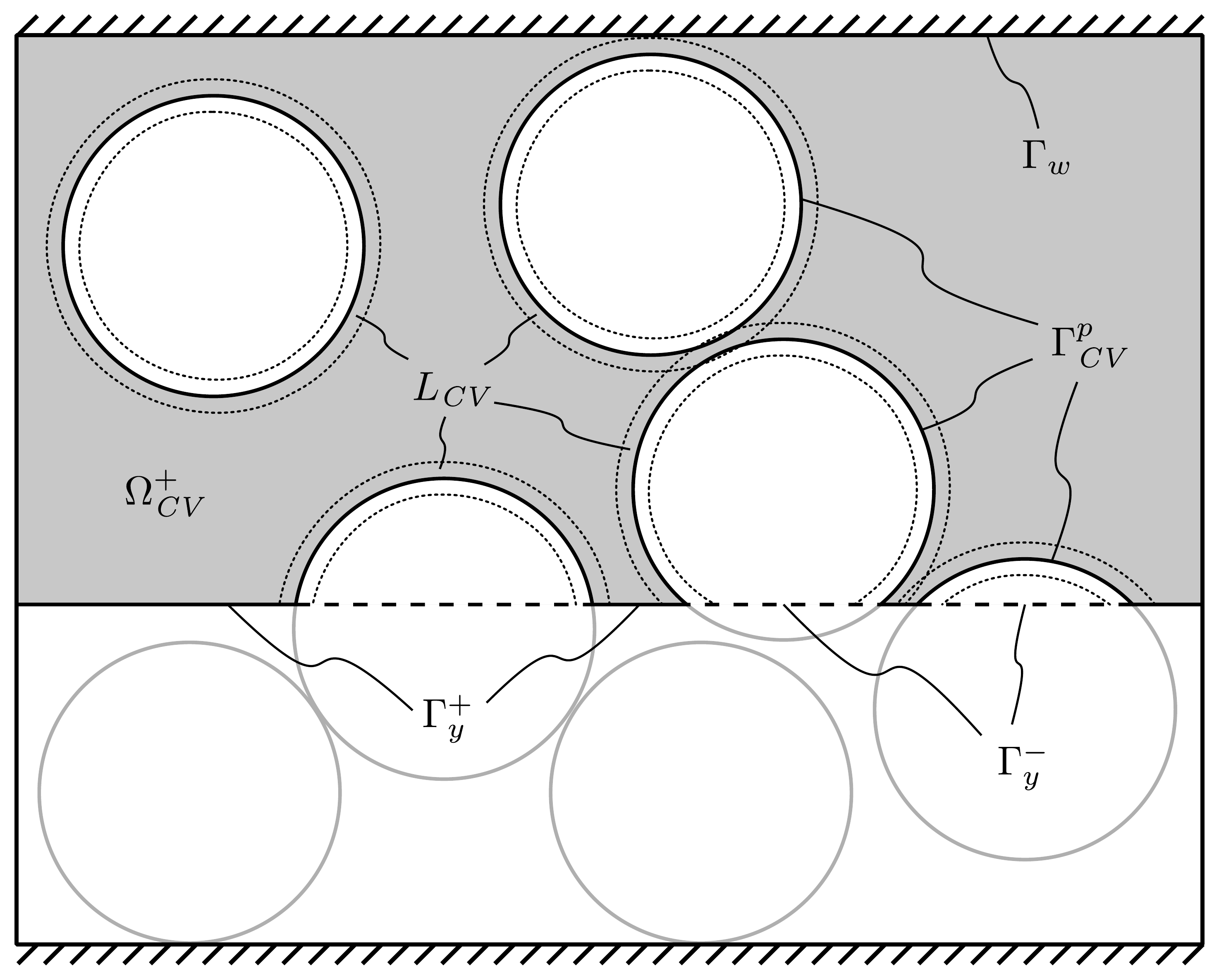}
\caption{Shaded control volume for the averaging involving multiple particles. All of the volumes and surfaces required for \eqref{eq:momx_fluid} are indicated.}  \label{fig:control_volume_general}
\end{figure}

To obtain the total shear stress, we consider the momentum balance 
\RII{
of the fluid phase, i.e. the Navier-Stokes equations \eqref{eq:NS},
}
over the control volume $\Omega_{CV}$ in the $x$-direction spanning from the top wall $\Gamma_w$ to some arbitrary height $y$ with the lower boundary being $\Gamma_y$ and involving multiple particles (figure~\ref{fig:control_volume_general}). We distinguish the free fluid from the particle interior as exemplified in figure \ref{fig:control_volume_general} using $\Omega_{CV}^+$ and $\Gamma_y^+$ for parts of the domain that are occupied by fluid and the lower boundary, respectively. 

\BV{
We can simplify the momentum balance of the fluid phase for the following reasons.
}
Owing to the periodic boundary conditions and the fact that we impose the Poiseuille flow via the volume force $\mbs{f}_b$, the pressure term ($\partial p_f / \partial x$ in \eqref{eq:NS}) does not contribute to the $x$-momentum, since our control volume covers the entire streamwise extent of the domain. At the top wall $\Gamma_w$, the vertical velocity, $v_f$, is zero, so that only $\eta_f \partial u_f/\partial y$ contributes to the fluid stress.  
\BV{
At the lower boundary, $\Gamma_y$, the pressure again does not play a role and the convective term vanishes due to the laminar flow conditions, so that the  long-range hydrodynamic stress originates from the viscous term.
}
Finally, we include $f_{IBM,x}$  as the sum of all hydrodynamic stresses arising from pressure and viscous forces that act on the particle surfaces $L_{CV}$ enclosed in $\Omega_{CV}$. These assumptions yield

\begin{IEEEeqnarray}{r} \label{eq:momx_fluid}
\underbrace{
\int\limits_{\Gamma_w} \eta_f \frac{\partial u_f}{\partial y} \,\mrm{d}A
+ \int\limits_{\Omega_\mathit{CV}} f_{b,x} \,\mrm{d}V
}_\text{External force}
= \underbrace{
\int\limits_{\Gamma_y^+} \eta_f \left(\frac{\partial u_f}{\partial y} + \frac{\partial v_f}{\partial x}\right) \,\mrm{d}A
}_\text{Fluid force} \nonumber\\
\underbrace{
- \int\limits_{L_\mathit{CV}} f_{\mathit{IBM,x}}\, {\mrm{d}V}
}_\text{Particle force} .
\end{IEEEeqnarray}
\BV{
The last term {\em Particle force} provides the linkage to the momentum balance of the particle phase, which can be obtained by appling the coarse-graining method \eqref{eq:cg_general} to the particle equation of motion \eqref{eq:p_translational}. This yields 
\begin{equation} \label{eq:p_int1}
\mbs{a}^{cg} = \mbs{F}_h^{cg} +  \mbs{F}_g^{cg} + \mbs{F}_i^{cg} ,
\end{equation}
where $\mbs{a}^{cg}$, $\mbs{F}_h^\mathit{cg}$, $\mbs{F}_\mathit{g}^\mathit{cg}$ $\mbs{F}_\mathit{i}^\mathit{cg}$ are the coarse grained forces due to the particle acceleration, hydrodynamic stress due to the IBM, gravity and particle interaction, respectively. Similar to the fluid momentum balance, we can analyze the coarse-grained particle forces within a control volume spanning the entire domain in the streamwise and spanwise directions and extending from the top wall to an arbitrary height $y$.  Integrating \eqref{eq:p_int1} over this volume, we obtain
\begin{equation} \label{eq:p_int2}
\int\limits_{\Omega_\mathit{CV}} \mbs{a}^{cg} \,\mrm{d}V = \int\limits_{\Omega_\mathit{CV}} \left(\mbs{F}_h^{cg} + \mbs{F}_g^{cg} + \mbs{F}_i^{cg} \right) \,\mrm{d}V .
\end{equation}
If particles are in a steady state, either naturally or through double-averaging, then the acceleration term vanishes. Furthermore, our setup of a horizontal channel yields $F_\mathit{g,x}=0$. We can, therefore, use the fact that $f_{IBM,x}=F_{h,x}^{cg}=-F_{i,x}^{cg}$ and apply \eqref{eq:particle_forces} to further distinguish between normal and tangential forces due to particle contact and friction ($F_{n,x}^{cg}$ and $F_{t,x}^{cg}$) and  short-range hydrodynamic lubrication forces ($F_{l,x}^{cg}$).
}

Using the averaging operator \eqref{eq:averaging_operator} and dividing by the horizontal area of the domain, we can rewrite \eqref{eq:momx_fluid} as
\begin{IEEEeqnarray}{r} \label{eq:fx_stress}
\underbrace{
\eta_f \overline{\left<\left.\frac{\partial u_f}{\partial y}\right|_{L_y}\right>}
\:+\: \overline{f}_{b,x} (L_y - y)
}_\text{External/Total stress}
= \underbrace{
\eta_f \overline{\left<\gamma\left.\left(\frac{\partial u_f}{\partial y} + \frac{\partial v_f}{\partial x} \right)\right|_y \right>}
-\int_y^{L_y} \overline{\left<F_{l,x}^{cg}\right>} \,\mrm{d}y
}_\text{Hydrodynamic stress} \nonumber\\
\underbrace{
+\: \int_y^{L_y} \left(\overline{\left<F_{\mathit{n,x}}\right>}+\overline{\left<F_{\mathit{t,x}}\right>}\right) {\mrm{d}y}
}_\text{Contact stress} ,
\end{IEEEeqnarray}
where $\gamma$ is an indicator function for the fluid ($\gamma = 1$ outside the particle and $\gamma = 0$ inside the particle), resulting in double-averaged equations for laminar flows akin to \citet{Nikora2013} and \citet{vowinckel2017a,vowinckel2017b}. We have also used the fact that $\eta_f$, $\rho_f$, and $f_{b,x}$ are constant throughout the domain. 
\BV{
It is important to note that we are explicitly separating the stresses arising from hydrodynamic interactions and particle contacts, respectively \citep{Gallier2014,gallier2018}, which will be analyzed in more detail in \S \ref{sec:rheology}.  The right-hand side of \eqref{eq:fx_stress} comprises the long-range fluid stress due to viscous effects evaluated at height $y$ at the boundaries of the control volumes $\Gamma^+_\mathit{CV}$ occupied by fluid as well as the short-range lubrication effects within the control volume $\Gamma^+_\mathit{CV}$. The contact stress comprises both, normal contact forces and tangential frictional forces.
} 
Note that it was shown in \citet{biegert2018a} and \citet{biegert2018b} that there is also a viscous and convective stress inside the particles $\Gamma^{-}_\mathit{CV}$ as a by-product of the IBM. For the present study, this effect does not contribute a significant part to the total stress balance. The external stress on the left-hand side of \eqref{eq:fx_stress} consists of the viscous stress at the top wall and the stress from the body force acting throughout the control volume. It is also equal to all other stresses arising from the movement of the fluid and particles and is hence equivalent to the total stress $\tau_f$ needed to compute the rheological quantities $\mu(J)$ and $\eta_s$. 

\begin{figure}
\placeTwoSubfigures{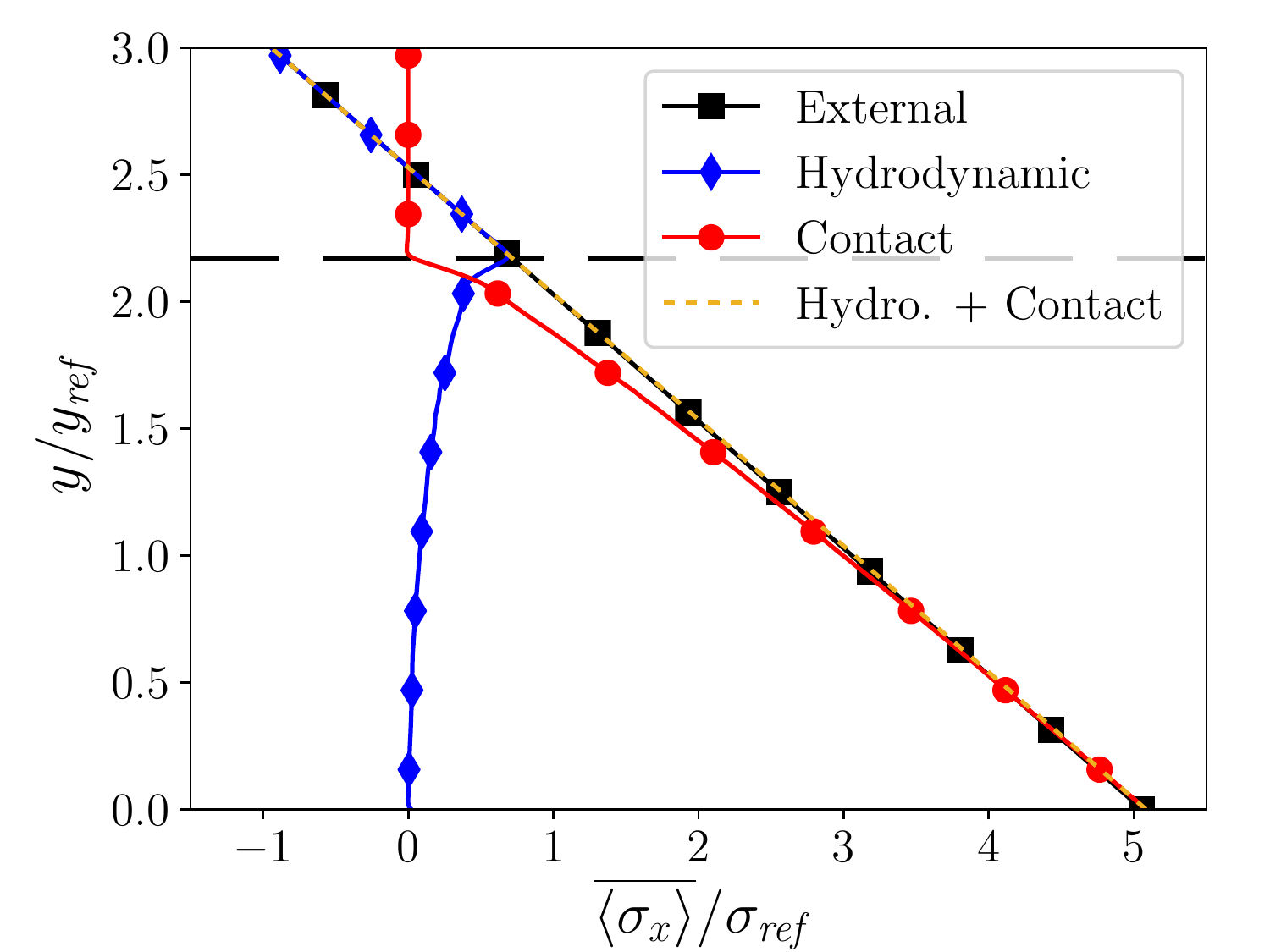}{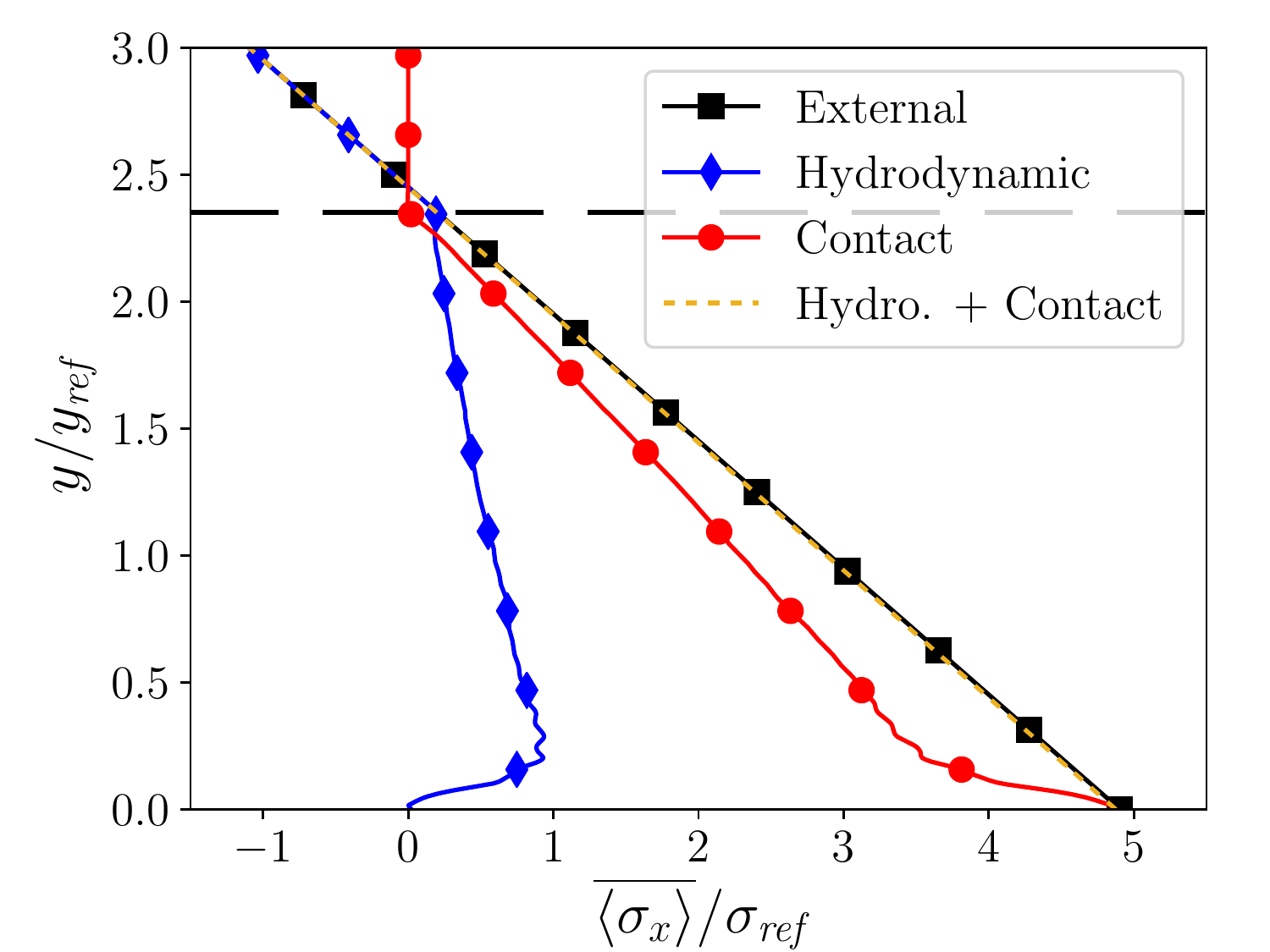}{0.75}{0pt}{0pt}{0pt}
\caption{Stress balance of the fluid phase in the $x$-direction  for a sheared particle bed according to \eqref{eq:fx_stress}. (a) run Re8 and (b) run Re33. The horizontal dashed line marks the height of the particle bed, $h_p$. As shown in (a) and (b), the sum of the hydrodynamic and contact stresses is in equilibrium with the external stress and consists of mostly the contact stress within the bed and the hydrodynamic stress above the bed.
}\label{fig:bed_momx_fluid}
\end{figure}

For the analysis of the stress balance of the fluid phase, we focus on runs Re8 and Re33 to get a sense of the results for different flow conditions. Figure~\ref{fig:bed_momx_fluid} shows the momentum balance of the fluid phase, given by \eqref{eq:fx_stress}, for runs Re8 (figure~\ref{fig:bed_momx_fluid}a) and Re33 (figure~\ref{fig:bed_momx_fluid}b), in which we expect the external stress to match the sum of the 
\BV{
hydrodynamic and contact stresses.
}
As expected, the external stress at the top wall is close to $\overline{\left<\sigma_x\right>}/\sigma_\mathit{ref}=-1$. In the upper part of the flow ($y/y_\mathit{ref} > 2.15$ and $y/y_\mathit{ref} > 2.3$ for Re8 and Re33, respectively), there are no particles, and the \BV{hydrodynamic} stress matches all of the external stress.  Within the particle bed ($y/y_\mathit{ref} < 2.3$), however, the majority of the external stress is taken up by \BV{particle contact}. The total stress comes out to be a linear profile. Hence, the present results support the conceptual model \eqref{eqqdmp2D} proposed by \cite{Aussillous2013} \RII{(figure 4b in this reference)} and we can use these data to compute $\eta_s$ and $\mu$ in the following.

Naturally, the \BV{hydrodynamic} stress of the laminar flow is entirely made up of the viscous term alone. Run Re8 differs from Re33 in that the \BV{hydrodynamic} stress reaches a higher positive value above the particle bed and quickly drops to zero within the particle bed.  The \BV{hydrodynamic} stress for Re33, on the other hand, 
\BV{
increases with increasing depth within the particle bed. Since the entire sediment bed is set in motion for this case, the lubrication component makes a significant contribution to the hydrodynamic stress.
}
These results are consistent with the velocity profiles in figure~\ref{fig:bed_flow}b, where the concavity of the profile for Re8 results in a high shear stress at the fluid/particle bed interface and low stresses within the bed, while the convexity of the profile for Re33 leads to a large shear stress at the lower wall. The \BV{total} stress, on the other hand, is completely dominated by \BV{particle contact within the sediment bed}. This result is consistent with the locations of the sharp gradients in the stress balance, so that the \BV{hydrodynamic and contact} stresses together close the $x$-momentum balance. It also shows that effects from the local acceleration term are negligible and that the unsteady flow conditions (figure \ref{fig:bed_p_flux}) are expected to have no impact on the reported results even for cases of dilating and consolidating sediment beds (e.g. Re33).

\subsection{Stress balance in the $y$-direction to obtain particle pressure}\label{sec:particle_pressure}
The rheological quantities $\eta_n$ and $\mu(J)$ require information about the particle pressure $p_p$, which we can obtain by further analyzing the particle phase.  Another way to interpret the bed particle pressure $p_p$  is to think of it as the total submerged weight of the particles \citep{vowinckel2019b}. Indeed, this has been done by \cite{Stickel2005} and \cite{Ouriemi2009a} for continuum modeling. We can demonstrate the validity of this reasoning by analyzing the momentum balance for the particle phase in the $y$-direction. 
\BV{
To this end, we utilize the coarse-grained momentum balance of the particle phase \eqref{eq:p_int1} and apply the averaging operator \eqref{eq:averaging_operator} to recast \eqref{eq:p_int2} as a line integral in the wall-normal direction
\begin{equation} \label{eq:p_int3}
\int_y^{L_y} \overline{\left<\mbs{a}^{cg}\right>} \,\mrm{d}y = \int_y^{L_y} \left(\overline{\left<\mbs{F}_h^{cg}\right>}  + \overline{\left<\mbs{F}_g^{cg}\right>} + \overline{\left<\mbs{F}_i^{cg}\right>} \right) \,\mrm{d}y .
\end{equation}
Since we are interested in the granular pressure, i.e. the bed weight, we consider only the vertical $y$-component. We again neglect the acceleration term, but keep the gravity term that is acting in the same vertical direction. Furthermore, we subdivide $\mbs{F}_i^{cg}$ into the short-range hydrodynamic lubrication and particle contact component, $\mbs{F}_l^{cg}$ and $\mbs{F}_n^{cg}$ and $\mbs{F}_t^{cg}$, respectively:
\begin{IEEEeqnarray}{r}  \label{eq:py_stress}
\underbrace{-\int_y^{L_y} \overline{\left<F_{g,y}^{cg}\right>} \,\mrm{d}y}_\text{Bed weight}
= \underbrace{\int_y^{L_y} \overline{\left<F_{h,y}^{cg}\right>} \,\mrm{d}y+\int_y^{L_y} \overline{\left<F_{l,y}^{cg}\right>} \,\mrm{d}y}_\text{Hydrodynamic stress}\nonumber\\
+ \underbrace{\int_y^{L_y} \overline{\left<F_{n,y}^{cg}\right>} \,\mrm{d}y+\int_y^{L_y} \overline{\left<F_{t,y}^{cg}\right>} \,\mrm{d}y}_\text{Contact stress} .
\end{IEEEeqnarray}
}
\RII{Since $\mbf{F}_{g,p}= V_p(\rho_p-\rho_f)\mbf{g}$, coarse-graining the particle weight yields the left-hand side as the submerged {\em bed weight}, which is equivalent to the particle pressure:}
\begin{equation}\label{eq:particle_pressure}
p_p = (\rho_p - \rho_f) \lvert \mbs{g} \rvert \int_y^{L_y} \overline{\left< \phi \right>} \, \mrm{d}y .
\end{equation}
\RII{We also note that \eqref{eq:particle_pressure} is the integral of \eqref{eqpp}, so that the present derivation provides a direct linkage to the two-phase modeling of \cite{Aussillous2013}. Owing to the fact that we obtain full information of the solid volume fraction (cf. figure \ref{fig:bed_flow}b), we do not need to introduce an artificial pressure at the fluid sediment interface. This was suggested by \cite{Houssais2016} (called $P_0$ in that study), but using such an artificial pressure would neither be in line with the two-phase modeling, nor would it obey the definition of our control volume sketched in figure \ref{fig:control_volume_general}. Hence, omitting $P_0$ naturally yields $J\rightarrow \infty$ for $\phi\rightarrow 0$, because $p_p\rightarrow 0$. This behavior is consistent with the approach of \cite{Boyer2011}. }

\begin{figure}
\placeTwoSubfigures{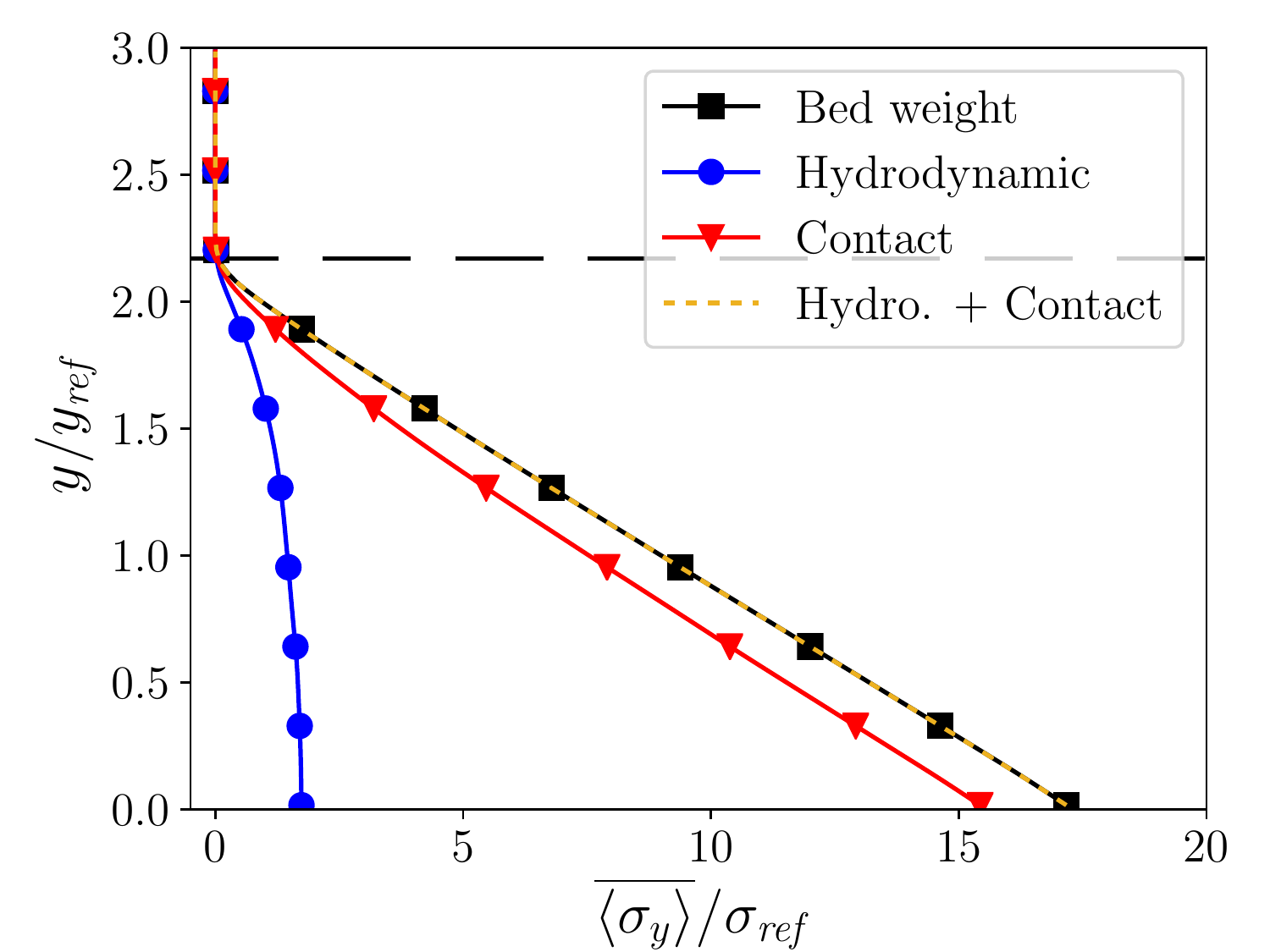}{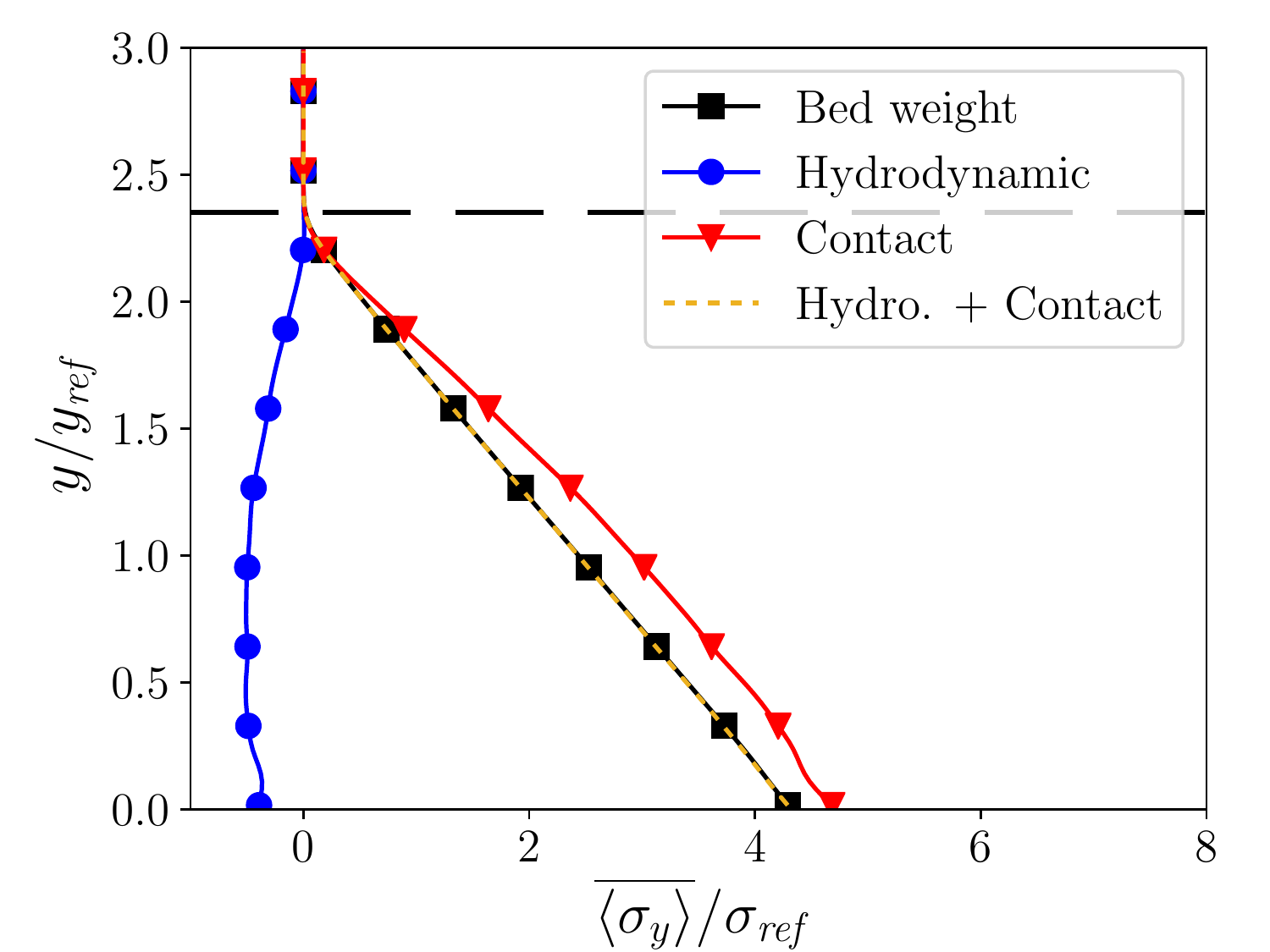}{0.75}{0pt}{0pt}{0pt}
\caption{Stress balance of the particle phase in the $y$-direction for sheared particle beds according to \eqref{eq:py_stress}. (a) run Re8 and (b) run Re33.  The horizontal dashed line marks the height of the particle bed, $h_p$. Frames (a) and (b) show that the sum of the hydrodynamic and contact stresses is in equilibrium with the bed weight,
\RII{i.e. $p_p$,} with most of the weight supported by the contact stress. The hydrodynamic lift force acting on the particles can be (a) positive or (b) negative. 
}
\label{fig:bed_momy_particle}
\end{figure}

Figure~\ref{fig:bed_momy_particle} shows the coarse-grained particle phase stresses, given by the time average of \eqref{eq:py_stress} for runs Re8 (left column) and Re33 (right column). In figures~\ref{fig:bed_momy_particle}a and~\ref{fig:bed_momy_particle}b, the bed weight increases almost linearly from the top of the particle bed down to the lower wall, balanced by the sum of the hydrodynamic and collision stresses. Again, this observation confirms the conceptual model of \eqref{eqpp} of \cite{Aussillous2013}. The fact that we have found a linear profile for this physical quantity simplifies the situation from a modeling perspective, i.e. one needs the sediment height and the total submerged weight of the sediment bed to reconstruct depth-resolved profiles of $p_p$. The results for the stress balance in $y$-direction show clear differences between runs Re8 and Re33. Owing to the normalisation, the dimensionless  $y$-momentum collision stress is three times larger for Re8 than for Re33. Moreover, the hydrodynamic stress is positive for Re8 and negative for Re33, so that the collision stresses are less than and greater than the bed weight, respectively, for the stress balance to be in equilibrium.  These deviations from zero stem from the fact that Re8 and Re33 are in the process of consolidation and dilation, respectively. 

\section{Rheology}\label{sec:rheology}
\begin{figure}
\begin{center}
\includegraphics[trim=3.5cm 0 4cm 0, clip,width=1\textwidth]{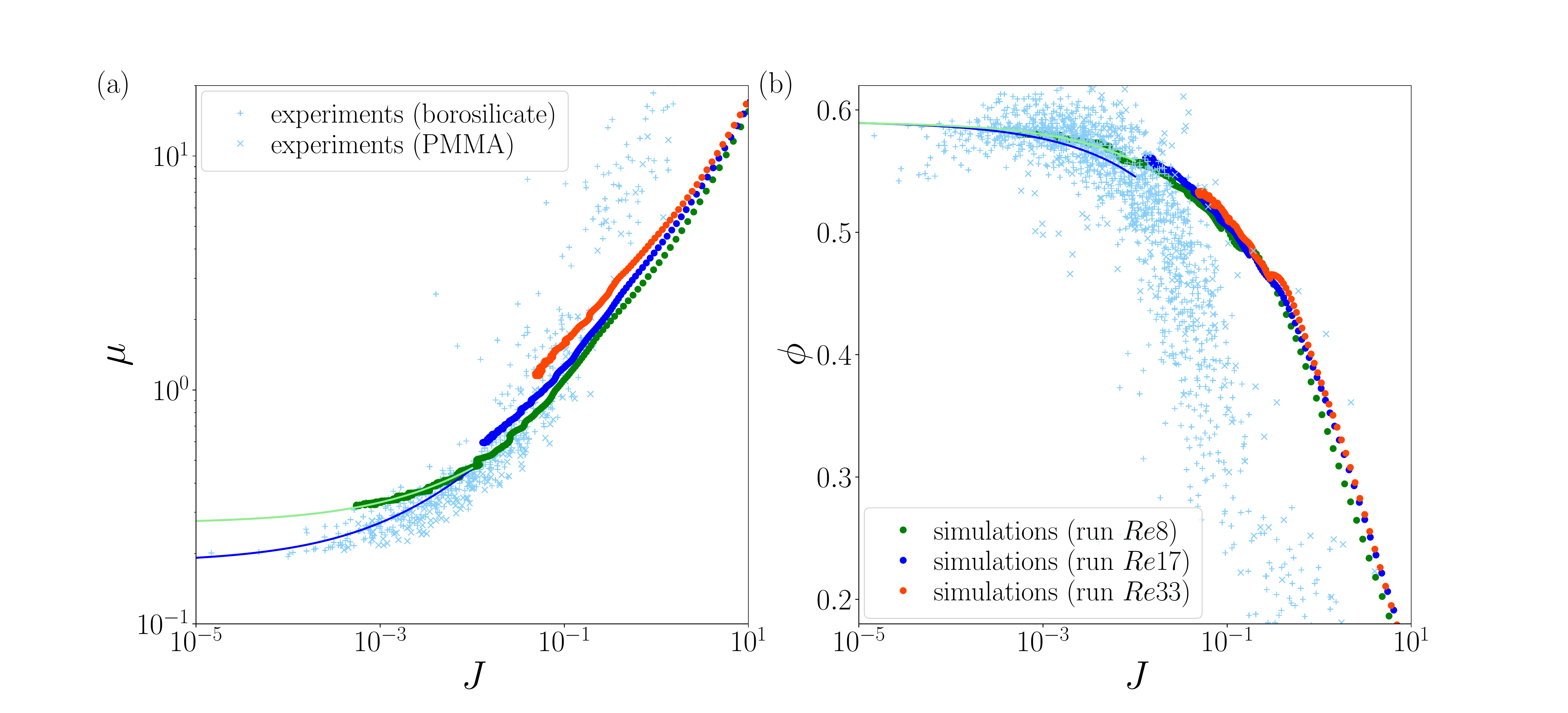}
\caption{(a) Effective friction coefficient, $\mu=\tau/p_p$, and (b) bulk volume fraction, $\phi$, versus the viscous number, $J=\eta_f \dot{\gamma}/p_p$. The solid lines are the asymptotic linear regressions in $\sqrt{J}$ (blue for the experiments and green for the simulations).}
\label{fig:muphivsJ_expsim} 
\end{center}
\end{figure}

We now turn to the examination of the rheological quantities extracted from the experimental measurements in \S\,\ref{sec:experimental_data} and from the numerical simulations in \S\,\ref{sec:simulations} and \ref{sec:stress_balance}. We display these data within the pressure-imposed view by plotting the effective friction coefficient, $\mu=\tau/p_p$, and the bulk volume fraction, $\phi$, as a function of the viscous number, $J=\eta_f \dot{\gamma}/p_p$, in figure \ref{fig:muphivsJ_expsim}. The macroscopic friction coefficient, $\mu$, is found to be an increasing function of the viscous number, $J$, whereas the volume fraction, $\phi$, is a decreasing function of the viscous number, $J$, for both, the experimental data coming from the two batches of spheres made of borosilicate ($+$) and PMMA ($\times$), \RII{where no appreciable differences can be seen between the runs of these two particle materials,} and the numerical data coming from the three different runs at different Reynolds numbers. 

It is important to note, however, that figure \ref{fig:muphivsJ_expsim} is not meant as a validation of the numerical results by comparing it to the post-processed experimental data of \cite{Aussillous2013} for the following reasons. First, the two data sets  represent different conditions. The particle properties $e_\text{dry}$ and $\mu_f$ chosen for the simulations may not correspond to the (unknown) values of the experiments and the flow conditions expressed by the Stokes number differ by more than one order of magnitude. Second, owing to the difficulty to  capture experimentally the sediment transport layer close to the bed interface of only a few particle diameters thickness, the experimental data present significant scatter. Finally, recall that our data processing in \S \ref{sec:experimental_data} assumes $u_f=u_{p,in}$ to reconstruct the experimental velocity profile, which slightly underestimates the fluid velocity in the sediment transport layer. On a similar note, the averaging operator using coarse-grained quantities \eqref{eq:cg_general} for the simulations yields an averaging window of three diameters in height to smooth out the sub-particle scale. For all these reasons, we can observe a rather large discrepancy between the experimental and numerical data for large $J$, i.e. in the sediment transport layer, but the qualitative trend is correctly reproduced. 

Despite the discrepancy at large $J$, both data sets show asymptotic behavior for small $J$ in figure \ref{fig:muphivsJ_expsim}b. The results can therefore be used to deduce the critical parameters $\mu_c$ and $\phi_c$ and provide proper scaling for our rheological analysis. Among the numerical results, only the data for the smallest Reynolds number (run Re8) reach the limit of small $J$, i.\,e. $J<10^{-2}$ and, hence, small $\dot{\gamma}$. The data coming from run Re8 and from the experiments show that both $\phi$ and $\mu$ tend to finite measurable values, $\phi_c$ and $\mu_c$ respectively, at vanishing $J$, i.\,e. at the jamming point where the suspension ceases to flow. The critical (or maximum flowable) volume fraction, $\phi_c$, and friction coefficient, $\mu_c$, can be measured by fitting the data using an asymptotic linear regression in $\sqrt{J}$ for small $J$ (for $J<10^{-2}$) as done by \cite{tapia2019}. The critical volume fraction is found to be $\phi_c\approx 0.59$ for both the experiments and simulations and is similar to the value found for suspensions made of spheres having only small surface roughnesses or equivalently having moderate inter-particle sliding friction coefficients \citep{Boyer2011,Dagois2015,tapia2019}. The critical friction coefficient is found to be $\mu_c \approx 0.18$ for the experiments and  $\mu_c\approx 0.27$ for the simulations. These values are smaller than those found previously ($\mu_c \approx 0.30-0.37$) in pressure-imposed rheometry \citep{Boyer2011,Dagois2015,tapia2019} but are of the same order of magnitude or even slightly larger than that given ($\mu_c \approx 0.17$) by the correlation of \cite{Morris1999} which was meant to match experimental results on shear-induced migration.

\begin{figure}
\begin{center}
\includegraphics[trim=3.5cm 0 4cm 0, clip,width=1\textwidth]{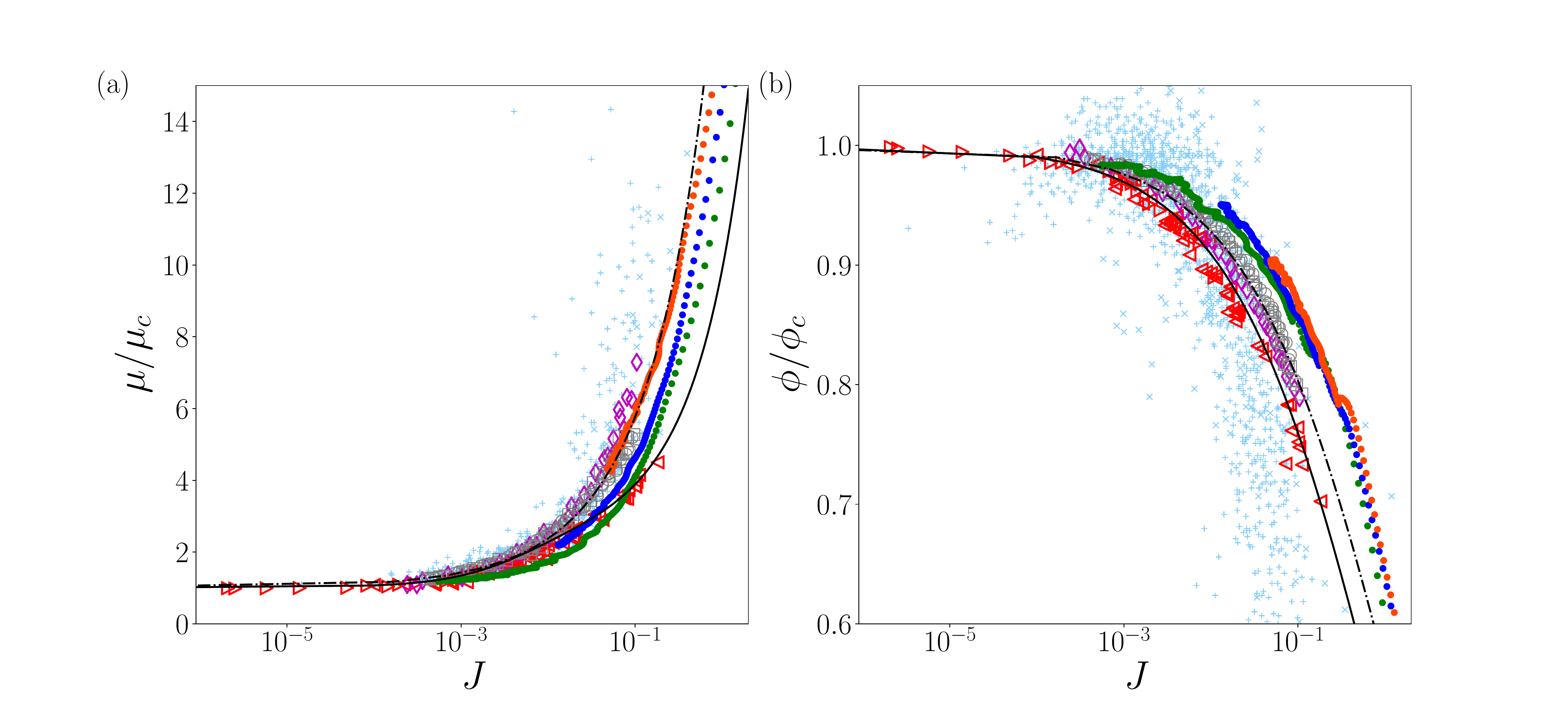}
\caption{Scaled (a) effective friction coefficient, $\mu/\mu_c$, and (b) bulk volume fraction, $\phi/\phi_c$, versus the viscous number, $J=\eta_f \dot{\gamma}/p_p$. Legend as in figure \ref{fig:muphivsJ_expsim}. Comparison with the experiments of \cite{Boyer2011} with polystyrene (PS: \textcolor{red}{$\boldsymbol{\triangleleft}$}) spheres of diameter $d_p=580 \, \mu$m suspended in polyethylene glycol-ran-propylene glycol monobutylether (PEG) as well as poly(methyl methacrylate) (PMMA: \textcolor{red}{$\boldsymbol{\triangleright}$}) spheres of diameter $d_p=1100 \, \mu$m suspended in a Triton X-100/water/zinc chloride mixture, of \cite{Dagois2015} with PS spheres of similar sizes suspended in PEG (\textcolor{purple}{$\boldsymbol{\Diamond}$}), and of \cite{tapia2019} with similar PS spheres which present small roughnesses (SR: \textcolor{gray}{$\boldsymbol{\circ}$}) and with PS spheres which are highly roughened (HR: \textcolor{gray}{$\boldsymbol{\square}$}). Comparison with the correlations proposed by \cite{Morris1999} (\sampleline{dash pattern=on .7em off .2em on .05em off .2em}) and \cite{Boyer2011} (\sampleline{}).
}
\label{fig:muphivsJ} 
\end{center}
\end{figure}

The critical values $\phi_c$ and $\mu_c$ are not universal parameters but depend on particle properties, i.\,e. the values of $\phi_c$ and $\mu_c$ depend on the particle size distribution but also on their surface properties. It is thus convenient to plot the data of figure \ref{fig:muphivsJ_expsim} by scaling $\phi$ by $\phi_c$ and $\mu$ by $\mu_c$ for a comparison with pressure-imposed rheological measurements \citep{Boyer2011,Dagois2015,tapia2019} and correlations \citep{Boyer2011,Morris1999}. These scaled data shown in figure \ref{fig:muphivsJ} collapse reasonably well onto the same constitutive curves for the small $J$-range ($J\lesssim10^{-2}$) but present discrepancies at larger $J$, i.\,e. low $\phi$. 
\RIII{However, the discrepancy at low $\phi$ is not surprising, because the empirical correlations of \cite {Boyer2011} and \cite{Morris1999} have been derived by fitting to experimental data of volume fractions $\phi/\phi_c>0.5$ for neutrally buoyant particles that are homogeneously distributed in a flow cell. In fact, it
}
is worth noting that there even exist some disparities within the pressure-imposed measurements, as while the data of \cite{Dagois2015} and \cite{tapia2019} are rather similar, they differ from the data of \cite{Boyer2011} at large $J$. This is reflected in the correlation of \cite{Boyer2011} which is designed to agree with the experimental measurements of \cite{Boyer2011}. 
This correlation seems to provide a lower bound for the whole collection of data while that of \cite{Morris1999} is more like an upper bound 
\RIII{for the data of $\mu(J)$ reported in figure \ref{fig:muphivsJ}a. The plotted data, therefore illustrates the range of uncertainty related to the extrapolation of empirical relations.}
It is important to stress that these two correlations have different critical values: $\phi_c=0.585$ and $\mu_c=0.32$ for the correlation of \cite{Boyer2011} and $\phi_c=0.68$ and $\mu_c=0.17$ for the correlation of \cite{Morris1999}. Note that value of $\phi_c=0.68$ is rather unrealistic as it exceeds the randomly closed packing fraction.

\begin{figure}
\begin{center}
\includegraphics[trim=3.5cm 0 4cm 0, clip, width=1\textwidth]{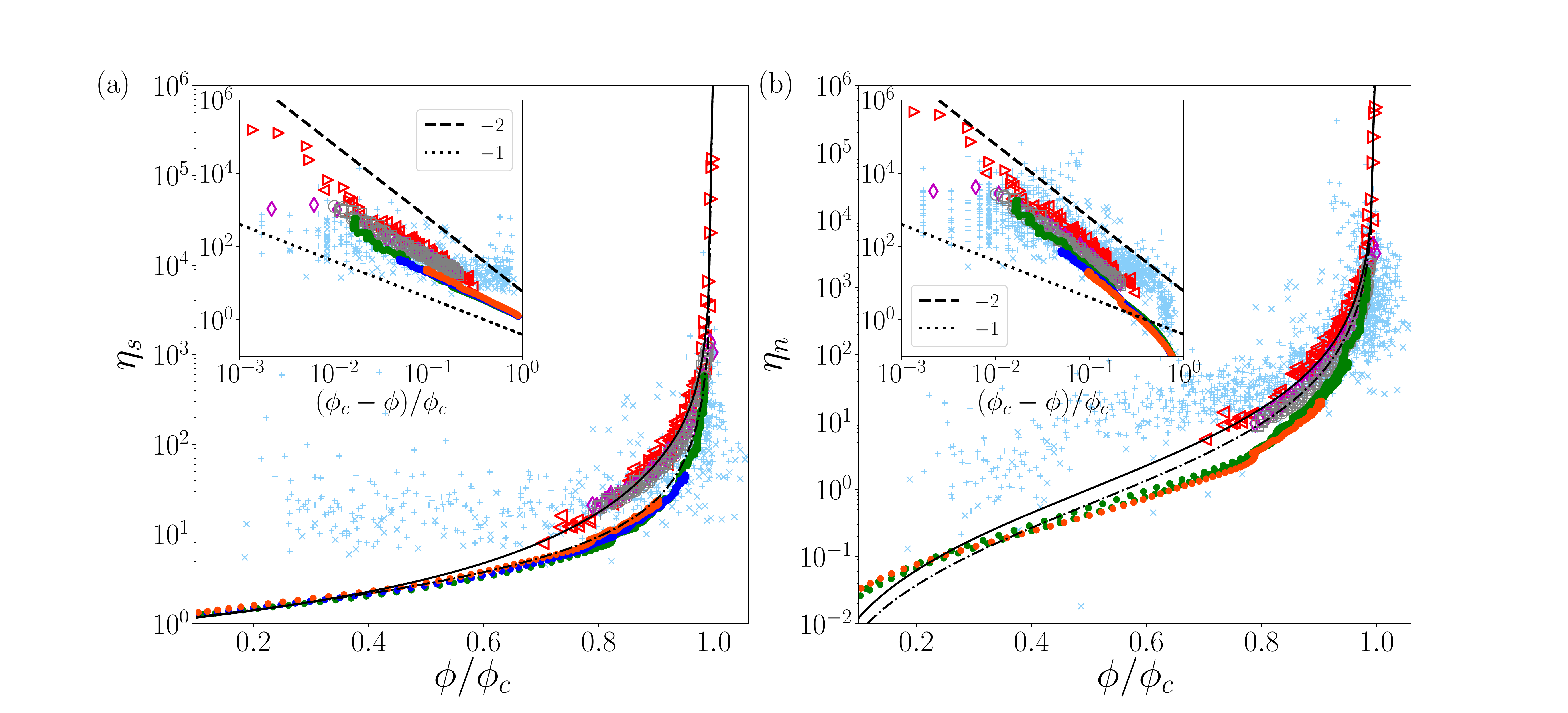}
\caption{(a) Shear, $\eta_s  = \tau/\eta_f\dot{\gamma}$, and (b) normal, $\eta_n =  p_p/\eta_f\dot{\gamma}$, viscosities versus the scaled volume fraction $\phi/\phi_c$. Same comparisons as in figure \ref{fig:muphivsJ}.}
\label{fig:etavsphi} 
\end{center}
\end{figure}

We also provide the classical volume-imposed view by plotting the shear, $\eta_s = \tau/\eta_f\dot{\gamma}=\mu/J$, and normal, $\eta_n=  p_p/\eta_f\dot{\gamma}=1/J$, viscosities as a function of the scaled volume fraction, $\phi/\phi_c$, in figure \ref{fig:etavsphi}. The collection of data shows that both viscosities increase with increasing $\phi$ and diverge at $\phi_c$.  The data coming from the erosion experiments follow this general trend despite the large scatters. Again, the correlations provide bounds for the whole collection of data, but this time the correlation of \cite{Boyer2011} acts as an upper bound while that of \cite{Morris1999} as a lower bound. The simulation data are notably lower than the pressure-imposed rheological measurements of \cite{Boyer2011}, \cite{Dagois2015}, and \cite{tapia2019} in particular for the smaller $\phi$-range. 

The log-log plots shown in the insets of figure \ref{fig:etavsphi} are made to analyze the asymptotic behaviors close to the jamming transition. The data for $\eta_n$ coming from run Re8  in the inset of figure \ref{fig:etavsphi} (b) present a divergence in $(1 - \phi/\phi_c)^{-2}$ in agreement with previous work \citep[see e.\,g.][]{Boyer2011,tapia2019}; the other runs are too far from the jamming point \RII{to} allow for any conclusions. The log-log plot shown in the inset of figure \ref{fig:etavsphi} (a) reveals that the same divergence dominates close to jamming for $\eta_s$. This last point is also demonstrated by the finite value of $\mu=\eta_s/\eta_n$ at vanishing $J$ for the data coming from run Re8 and from the erosion experiments in figure \ref{fig:muphivsJ_expsim} (a). 

Apart from the difficult task to accurately measure within the sediment transport layer of only a couple particle diameters thickness, the issues described above to recover the rheological quantities at large $J$ and small $\phi$ can be attributed to the fact that sediment transport can be sub-divided into two different regimens \citep{Revil-Baudard2015}. On the one hand, the dynamics of dense suspensions with large $\phi$ are dominated by frictional contact. In fact, this is the regime that has been investigated in the experiments of \cite{Boyer2011}, \cite{Dagois2015}, and \cite{tapia2019} and for which the empirical correlations shown in figures \ref{fig:muphivsJ} and \ref{fig:etavsphi} have been derived. On the other hand, low volume fractions represent the dilute regime of mostly binary collisions between particles. It was therefore concluded by \cite{Maurin2016} that those empirical correlations may need adjustments for these flow conditions. 

Hence, our observations are in line with the study of \cite{Houssais2016}, who conducted experiments of sediment transported in an annular flume and measured depth-resolved profiles very similar to the data presented here. In this study, the authors had to introduce a confinement pressure to correct $\mu$ for  large values of $J$. This pressure was added to the granular pressure and can be interpreted as the artificial weight of the top plate in a pressure-imposed rheometer. \cite{Houssais2016} found good agreement with the correlations of \cite{Boyer2011} for $\mu(J)$ using this measure, but unfortunately, the agreement for the correlation $\phi(J)$ was not as good in that study. 
\RII{As already mentioned in \S \ref{sec:particle_pressure}, we omit using this strategy of an additional confinement pressure in the present  analysis. The high resolution of the numerical data  did not require such a treatment, because all relevant quantities could be recovered by our statistical analysis in a straightforward manner. This way, we ensure to stay consistent with the two-phase flow model \eqref{eqqdmp2D} and \eqref{eqpp} from \cite{Aussillous2013}. The same applies for the revisited data of \cite{Aussillous2013}, but for these data we use the additional assumption of  ${u_f=u_{p,in}}$ to reconstruct the fluid velocity profile from a mixed Couette-Poiseuille-flow, which yields a slight underestimation of the fluid velocity in the range of large $J$ and low $\phi$. }

\begin{figure}
\centering
\includegraphics[width=0.8\textwidth]{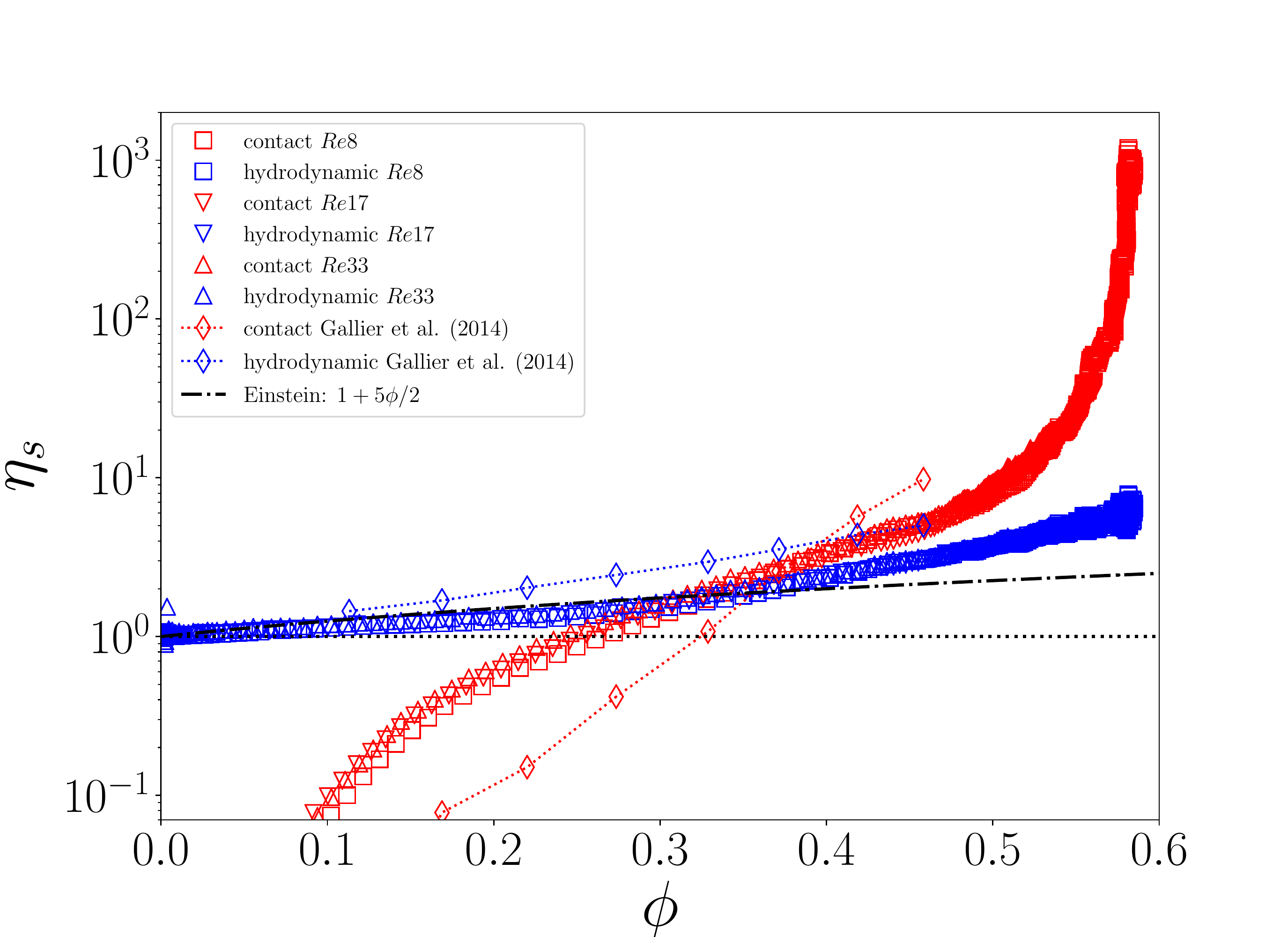}
\caption{Relative contributions of the frictional contact (red) and the hydrodynamic (blue) stress to $\eta_s$.}  \label{fig:eta_s_hydro_contact}
\end{figure}
\BV{
To further investigate the governing effects in the dilute and dense regime as a function of the particle volume fraction, we make use of the highly-resolved data and the stress balance \eqref{eq:fx_stress} derived in \S \ref{sec:stress_balance} and separate the stresses due contact and hydrodynamic interaction. These are plotted in figure \ref{fig:eta_s_hydro_contact} (without the normalization by $\phi_c$) together with a comparison of the data of \cite{Gallier2014}, who carried out three-dimensional simulations of neutrally buoyant, non-Brownian, frictional spheres in a Couette cell of constant size, i.e. volume-imposed rheometry. For our simulation results, the data of the three simulation runs collapse onto a single master curve for both the contact and the hydrodynamic stress components. For $\phi>0.3$, there are no significant long-range hydrodynamic effects possible and the particle contact is the main contribution to the total stress, whereas hydrodynamic effects prevail below this threshold value of $\phi$. The analysis therefore reveals the reason why the considerations for pressure-imposed rheometry also hold for sediment transport in the dense regime, i.e. $\phi>0.3$, where the rheology is dominated by particle contact. 
}

\BV{
In the dilute regime, i.e. $\phi<0.3$,  the hydrodynamic component scales very well with the Einstein relation $\eta_s = 1 + \frac{5}{2}\phi$ \citep{Einstein1956} and the contribution from particle contact becomes negligible. Our analysis furthermore shows that the deviation of the hydrodynamic component from its clear fluid value ($\eta_s=1$) is induced by the lubrication forces (not shown here). This observation differs from the simulation results of \cite{Gallier2014} obtained using volume-imposed rheometry for $0.1\leq \phi \leq 0.45$, where particles are distributed homogeneously in the simulation domain. In the present simulations, the dilute regime represents the thin layer of active sediment transport, where particles are still touching and aligned in the shearing direction so that the particle volume fraction decreases rapidly from densely packed to zero within a thin layer that is only a couple of diameters thick. Hence, long-range hydrodynamic effects are screened by the highly anisotropic distribution and the steep gradient of $\phi$ in this layer. As a result, the hydrodynamic component resembles a porous media flow behavior as anticipated in the two-phase modelling of \cite{Ouriemi2009a} and \cite{Aussillous2013}. 
}

To the knowledge of the authors, the only numerical study of particle-resolved DNS that was able to perform the analysis of the rheological behavior of a sheared sediment bed has been presented  by \cite{Kidanemariam2017Diss}. Interestingly, this study underestimates the empirical correlation for $\phi(J)/\phi_c$ for large $J$, whereas our simulation data overestimates the volume fraction in this regime (cf. figure \ref{fig:muphivsJ} (b)).
\RII{
The low values of $\phi$ for large $J$ in the data of \cite{Kidanemariam2017Diss} can in parts be attributed to the difference in handling particle collisions and contact. \cite{Kidanemariam2017Diss} did not resolve lubrication forces which imposes the constraint that one has to keep a minimal distance of two grid cells between particles.}
Consequently, the granular packing has to be less dense so that unusually low values of $0.43 \leq \phi_c \leq 0.53$ were reported for the maximum packing fraction. Hence, the data and the analysis presented here demonstrate a substantial improvement over previous efforts to capture the rheology of sediment beds over a wide range of $J$.

The rheological data of figures\,\ref{fig:muphivsJ_expsim}, \ref{fig:muphivsJ}, \ref{fig:etavsphi} and \ref{fig:eta_s_hydro_contact} are given as supplementary materials.

\section{Concluding remarks}\label{sec:conclusion}
\RI{Serving the need to use constitutive laws for macroscopic sediment transport models,}
the present paper introduced a new means to generate highly resolved data for the rheological behavior of granular sediment beds sheared by a viscous, pressure-driven  flow. The rheology is assessed by a statistical analysis to extract the physical quantities needed to compute the rheological parameters. 
\RI{The highly-resolved simulations provide the data needed to assess the stress exchange of the fluid-particle mixture by deriving the stress balance from first principles.} 
\RIII{To better understand naturally occurring sediment-laden flows in pipes and rivers, we focus on pressure-driven flows, where the total stress increases with flow depth.}
The analysis verified the conceptual two-phase model of \cite{Ouriemi2009a} and \cite{Aussillous2013} that has been derived from the Brinkman equations, and deduces the total fluid shear and the granular pressure as the relevant rheological quantities. We compare our numerical results to the rheology of corresponding experiments by revisiting the data set of
\RIII{pressure-driven flows investigated by \cite{Aussillous2013} and found reasonably good agreement between the numerical and experimental data.}
The results 
\RI{presented here clearly show that sediment transport by pressure-driven flows yields results that are similar to those of}
previous studies of annular Couette-type flows, 
\RI{even though these studies were conducted} 
with dense suspensions of neutrally buoyant particles in pressure-imposed rheometers. \RIII{The analyzed data agree well with the empirical  correlations of \cite{Boyer2011} and \cite{Morris1999} derived therefrom. In the more dilute regime, the simulation data fall in between the bounds set by those two extrapolated correlations. The good agreement of the data obtained from pressure driven flows with those from Couette flows} 
potentially justifies the use of these empirical correlations 
\RIII{ as constitutive equations} 
for two-phase flow solvers for sediment transport applications in the dense regime. 
\BV{Separating the total shear stress of our simulation results into the parts from hydrodynamic interaction and particle contact revealed a critical particle volume fraction of $\phi\approx0.3$, for which particle contact becomes the dominant component, so that classical empirical relations from rheometry become applicable. In the dilute regime, we obtain a scaling in agreement with the Einstein relation, which illustrates a flow behavior that differs from classical rheometry studies, because the long-range interactions are screened by the porous media. This effect corresponds to the thin layer of sediment transport over which hydrodynamic and contact stresses but also $\phi$ vary rapidly, which is very difficult to grasp, but it constitutes the main difference between classical rheometry of neutrally buoyant suspensions and the rheology of sediment transport. }    \RIII{More work of this kind will be needed in the future to address different particle properties and flow situations that are more complex than the two fundamental flow types Couette and Poiseuille flow.}

\section*{Acknowledgements}
B.V. gratefully acknowledges the Feodor-Lynen scholarship provided by the Alexander von Humboldt foundation (Germany) and the German Research Foundation (DFG) grant VO2413/2-1. Furthermore, this research was supported in part by the Department of Energy Office of Science Graduate Fellowship Program (DOE SCGF), made possible in part by  the American Recovery and Reinvestment Act of 2009, administered by ORISE-ORAU under contract no. DE-AC05-06OR23100.  It is also supported by the Petroleum Research Fund, administered by the American Chemical Society, grant number 54948-ND9.   E.M. gratefully acknowledges support through NSF grants CBET-1803380 and OCE-1924655, as well as by the Army Research Office through grant W911NF-18-1-0379, and from Petrobras.  Computational resources for this work used the Extreme Science and Engineering Discovery Environment (XSEDE), which is supported by the National Science Foundation, USA, Grant No. ACI-1548562. This work was supported by the Labex MEC (ANR-10-LABX-0092) under the A*MIDEX project (ANR-11-IDEX-0001-02) funded by the French government program Investissements d'Avenir.

\section*{Declaration of Interests} 
The authors report no conflict of interest.

\bibliographystyle{jfm}
\bibliography{jfm-instructions}

\end{document}